\documentclass[letterpaper]{article} % DO NOT CHANGE THIS
\usepackage{aaai2026}  % DO NOT CHANGE THIS (de-anonymizes if you delete [submission] 
\usepackage{times}  % DO NOT CHANGE THIS
\usepackage{helvet}  % DO NOT CHANGE THIS
\usepackage{courier}  % DO NOT CHANGE THIS
\usepackage[hyphens]{url}  % DO NOT CHANGE THIS
\usepackage{graphicx} % DO NOT CHANGE THIS
\usepackage{natbib}  % DO NOT CHANGE THIS AND DO NOT ADD ANY OPTIONS TO IT
\usepackage{caption} % DO NOT CHANGE THIS AND DO NOT ADD ANY OPTIONS TO IT
\usepackage{booktabs, multirow}
\usepackage[shortlabels]{enumitem}
\title{Spotting (and Missing) Algorithmic Bias: Investigating User Understanding in a Fairness Assessment Tool}

\author{
    Anna Verheyden\textsuperscript{\rm 1,\rm 2},
    Yizhe Zhang\textsuperscript{\rm 1,\rm 2},
    Robin De Croon\textsuperscript{\rm 1,\rm 2},
    Simone Stumpf\equalcontrib\textsuperscript{\rm 3},
    Katrien Verbert\equalcontrib\textsuperscript{\rm 1,\rm 2},
}

\affiliations {
    \textsuperscript{\rm 1} KU Leuven, Department of Computer Science, B-3001 Leuven, Belgium,\\
    \textsuperscript{\rm 2} Augment, imec research group at KU Leuven,\\
    \textsuperscript{\rm 3} University of Glasgow, Glasgow, Scotland, UK\\
    \{anna.verheyden, yizhe.zhang, robin.decroon, katrien.verbert\}@kuleuven.be,
    simone.stumpf@glasgow.ac.uk
}

\begin{document}

\maketitle

\begin{abstract}
Fairness metric selection is typically left to data scientists, but which biases are problematic and which metric captures them best depends on stakeholders' experience and domain knowledge. This calls for involving non-technical stakeholders, but the research prototypes built for this purpose so far have not tested whether these stakeholders form accurate mental models of the metrics they interact with or can act on them to identify biases.
We present FairAware, a fairness assessment tool co-designed with Human Resources (HR) domain experts. We evaluate stakeholders' understanding through a mixed-methods study with 70 participants (35 HR employees, 35 job seekers), measuring objective and subjective understanding, cognitive load, bias identification accuracy, and open-ended feedback. Most participants correctly identified the most disadvantaged group, with task duration being the only significant predictor. We also found a gap between subjective and objective understanding, with both groups performing similarly across all measures.  
These results suggest that fairness assessment tools for non-experts are usable for identifying biases but need built-in checks on understanding before stakeholders make higher-stakes decisions. 
 
\end{abstract}

% DONT FORGET ANONYMISATION
%\begin{links}
%    \link{Prototype}{https://fairness-dashboard-872410478845.europe-west1.run.app/?nostudy=1}
%\end{links}

\section{Introduction}
Recommendation algorithms are increasingly used to support high-stakes decisions in domains such as hiring \cite{Barocas2019}, which strongly affect people's lives. In the hiring domain, algorithmic tools are used to screen resumes, rank candidates, and recommend job matches \cite{Kchling2020}. However, this raises concerns around algorithmic bias: if the data or model encodes historical patterns of discrimination, the algorithm may systematically disadvantage certain demographic groups \cite{barocas2016_disp-imp}. Hence, there are ethical and legal reasons to ensure that they are as fair as possible. The EU AI Act \cite{EUAIAct2024} now requires human oversight of high-risk AI systems, which means companies are forced to pay attention to assessing the fairness of these systems.

Current ways of assessing fairness rely on fairness metrics.  More than fifty metrics have been proposed, many of which are mutually incompatible \cite{Verma2018, Kleinberg2017}, and there is no single correct way to assess algorithmic fairness. Selecting which metrics to apply is itself a value judgment that depends on domain context \cite{Cheng2021_FEI, Lee2019_webuildai}. Yet, the stakeholders that are best positioned to make that judgment due to their domain expertise or lived experience--HR employees and job seekers--usually lack the technical background to interpret the available metrics. 

This raises the question of whether interfaces designed to present fairness metrics to non-technical stakeholders support them in interpreting these metrics correctly and identifying potential algorithmic biases. While recent work has developed fairness tools for eliciting non-expert users' fairness preferences \cite{Cheng2021_FEI, Nakao2023, Luo2025_EARN}, or proposed participatory frameworks that include stakeholders in fairness-related decisions \cite{Cheng2021_FEI, Lee2019_webuildai}, the question of these stakeholders' understanding of it or their ability to find algorithmic biases has not been investigated.

To fill this gap, we present FairAware, a fairness assessment dashboard to evaluate recommendations for matching job vacancies, co-designed with non-technical stakeholders in the hiring domain, and a mixed-methods study with 35 HR employees and 35 job seekers exploring two research questions:
\begin{description}
    \item (RQ1) What affects users' subjective and objective understanding of fairness metrics, and their ability to identify biases?
    \item (RQ2) How do users employ the dashboard to advance their understanding and find biases? 
\end{description}
  
This work contributes: 
\begin{itemize}
    \item an artifact: FairAware, a fairness assessment dashboard co-designed with HR domain experts that lets non-technical users navigate four fairness metrics, weigh them according to their preferences, and drill down to details or a causal view for attribute relations.
    \item empirical findings: A mixed-methods study with 70 non-technical participants showed that most could correctly identify the most disadvantaged group, with time on task as the only significant predictor. Using a new \textit{mental model} instrument we constructed, we found that subjective understanding tracked prior familiarity rather than objective understanding, and dashboard interaction logs did not predict either, suggesting a misalignment between subjective and objective understanding and that participants spotted biases without necessarily understanding the underlying metrics.
\end{itemize}

\section{Background and Related Work}
\label{sec:related_work}
\subsection{Measuring Fairness and Involving Stakeholders to Assess Fairness}
There is no single, overarching metric for algorithmic fairness. Instead, over fifty metrics have been proposed, each one reflecting a different ethical viewpoint \cite{Verma2018}. 
These fairness metrics are mutually incompatible \cite{Kleinberg2017}, making metric selection a context-dependent value judgment \cite{Waller2025, Luoetal2026_ithinkthisisfair}. 
Group-level metrics (e.g., statistical parity, disparate impact) compare outcomes between a privileged and an unprivileged group defined by a sensitive attribute, while individual-level metrics (e.g., consistency, counterfactual fairness) ask whether similar people are treated similarly \cite{Verma2018, Kearns2018}.
Applying any of these metrics requires two choices that are usually made by data scientists rather than the affected stakeholders: which metric to choose, and which attribute defines the groups.
Most existing toolkits operationalize this for data scientists, for example, Fairlearn \cite{Bird2020}, Aequitas \cite{saleiro2019aequitasbiasfairnessaudit}, AI Fairness 360 \cite{Bellamy2019}, the What-If Tool \cite{Wexler2020}, or offer a specialized functionality, for example, FairVis \cite{Cabrera2019} addressing intersectional bias, or Silva \cite{Yan2020} and D-BIAS \cite{Ghai2023} visualizing causal relationships between attributes.

Because fairness is context-dependent and value-laden, recent research has started to involve non-technical stakeholders, either by eliciting their fairness notions without pre-defining the metrics \cite{nakao_toward_2022}, or by surfacing their varying preferences once metrics are explained \cite{Cheng2021_FEI, Lee2019_webuildai, Luoetal2026_ithinkthisisfair}. 
Several research prototypes have been built to support non-technical stakeholders in assessing fairness: FairHIL \cite{Nakao2023}, the EARN Fairness Framework with the FEE tool \cite{Luo2025_EARN}, and the Fairness Elicitation Interface \cite{Cheng2021_FEI}, which focus on surfacing harms, eliciting preferences, or facilitating collective deliberation. 
However, these prototypes assume that stakeholders understand the metrics they interact with, an assumption that has not been tested. If their comprehension is inaccurate, the value of their input is undermined. 
They have also been evaluated on loan application (FairHIL, FEE) or child welfare prediction (FEI), but not in recruitment, where stakeholder needs and fairness preferences might differ. 

Our work takes inspiration from the above studies by building an interface that targets non-technical stakeholders, but differs from these approaches in two ways: (1) we focus on measuring whether non-technical stakeholders actually understand, and can act on, the fairness metrics and the tool presented to them, and (2) we do so in the recruitment domain, a domain where the EU AI Act applies directly. For (1), we validate their objective understanding, measure whether it aligns with their subjective understanding, and assess its relation to cognitive load. We involve stakeholders throughout the full process, which introduces specific stakeholder needs.

\subsection{Measuring Understanding}
Users' subjective and objective understanding of AI systems and tools is frequently assessed in eXplainable AI (XAI). \citeauthor{Hoffman2023} provides a comprehensive overview of methods to do so, including ways to measure users' mental models. A number of question types have been developed to elicit and measure mental models \cite{Kulesza2012, Lakkaraju2016, Bhattacharya2024, Bhattacharya2025, Nimmo2024}: 1) fact questions that test retention of specific terminology, 2) simulation or prediction questions that ask users to anticipate system behavior, and 3) articulation questions that require free-text explanations. Often, multiple-choice answers are used in response to these questions, but this format risks overestimating objective understanding because participants can recognize an answer without being able to reproduce it, or they can simply guess \cite{cueingeffectsmcq}. 

No validated instruments exist for measuring mental models of fairness metrics specifically. Existing questionnaires target model explanations (e.g., feature attributions, decision rules), vary widely in scoring and item composition, and are very context-specific; no shared guidelines for mental model questionnaire construction exist to our knowledge. Our work contributes the construction process of a new instrument, balancing fact, simulation, and articulation questions to capture understanding at multiple depths, based on prior mental-model studies in XAI. The design choices, scoring, and trade-offs are described in the Methods section. 

\section{Methods}
\label{sec:methods}
Following a human-centered co-design approach (n=15), we first designed a prototype, the FairAware dashboard, to support non-technical stakeholders in assessing fairness and identifying biases. We then conducted a mixed-methods online study with 70 participants (35 HR employees and 35 job seekers) to investigate our main research questions. All research received ethical approval from the KU Leuven Ethics Committee with an approval number G-2025-9565-R2(AMD), and all participants provided informed consent.

\subsection{Prototype: The FairAware Dashboard}
\label{subsec:prototype}
\begin{figure*}
    \centering    \includegraphics[width=\linewidth]{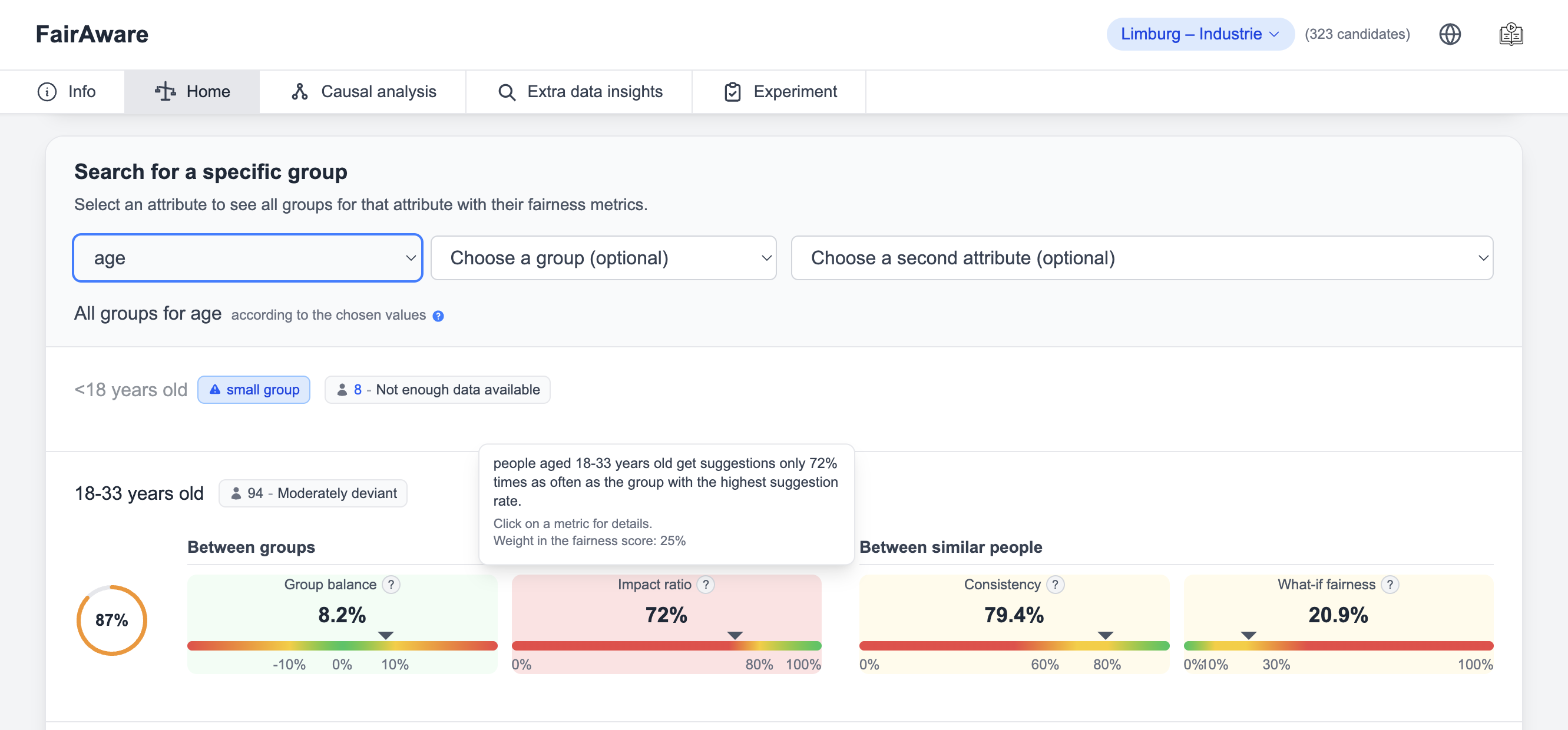}
    \caption{The \textit{Home Page} of FairAware, more specifically, the component to search the results for a specific attribute. Here, age is chosen, and the results are visible for the group 18-33 years old. The 87\% on the left is the `overall fairness score', which is the weighted sum of the four metrics to the right of it (after rescaling), based on the participant's chosen metric priorities.}
    \label{fig:dashboard}
\end{figure*}
\begin{figure}
    \centering
    \includegraphics[width=\linewidth]{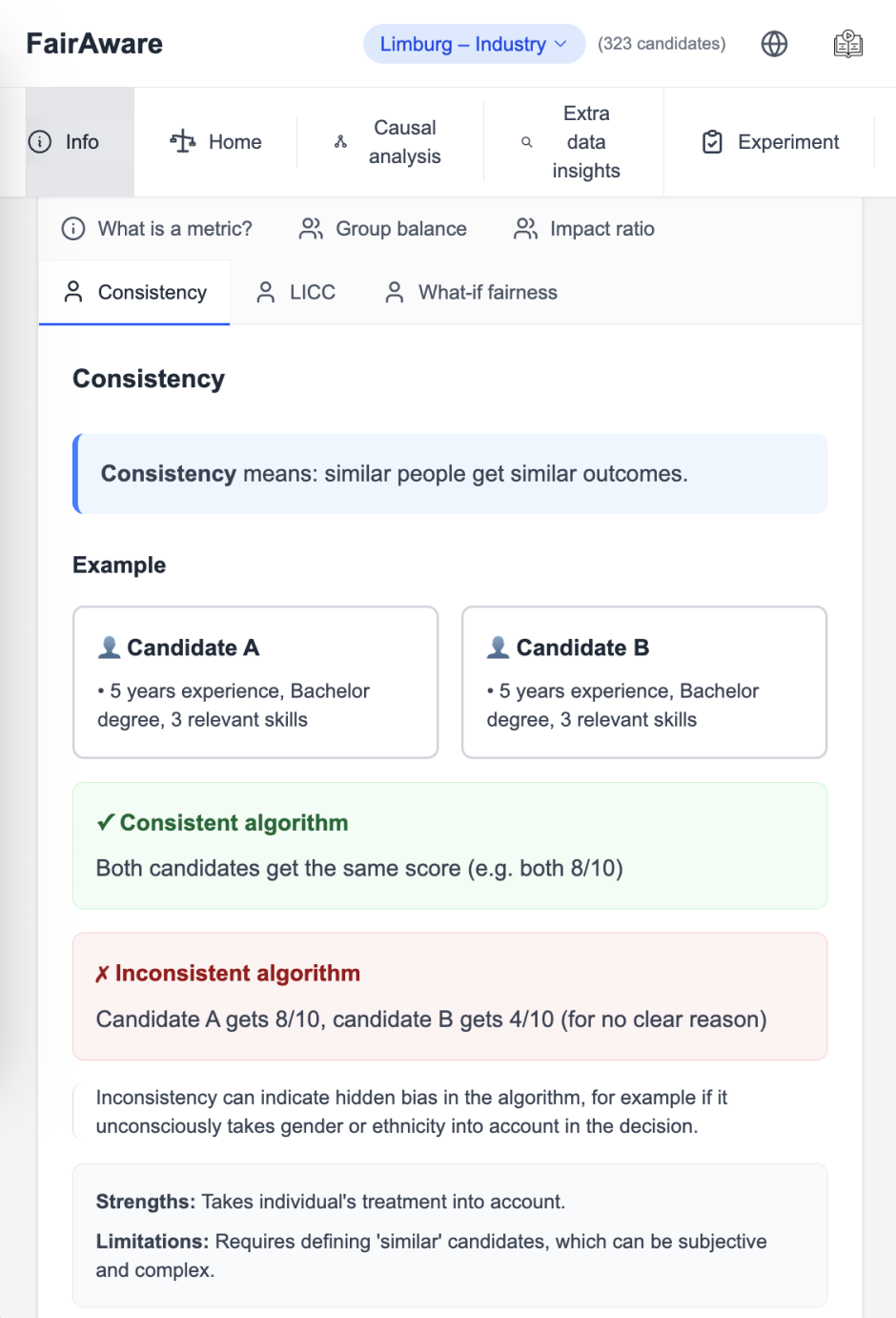}
    \caption{The \textit{Information Page} of FairAware. On top, the different metrics can be selected to see their explanation. Currently, consistency is selected. The one-sentence explanation is shown, followed by an example, extra relevant info, a strength, and a limitation of the metric. }
    \label{fig:infopage}
\end{figure}
\begin{figure}
    \centering
    \includegraphics[width=\linewidth]{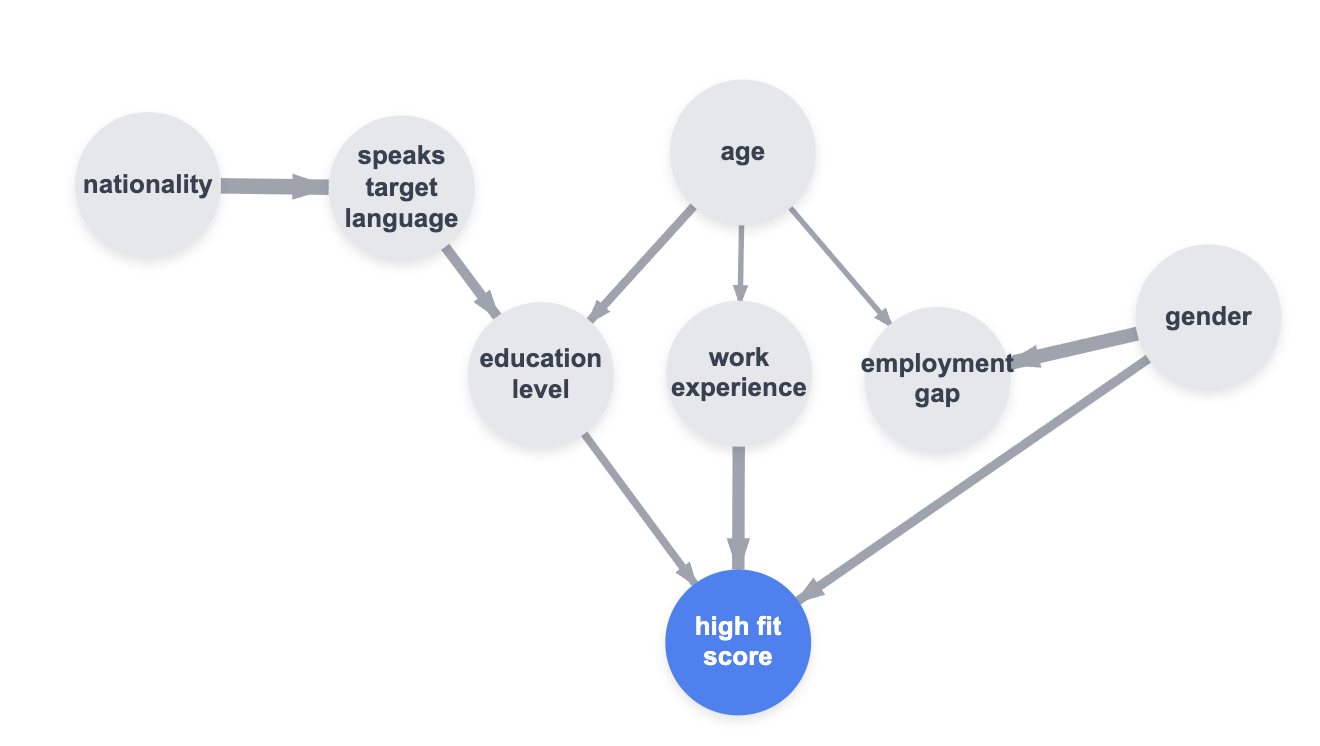}
    \caption{The causal graph of the \textit{Causal Analysis Page} of FairAware. Nodes show the attributes, edges show the correlations, with edge thickness indicating the strength of the correlation. High fit score indicates the similarity score between a vacancy and a candidate, which is used to decide whether they get a vacancy suggested.}
    \label{fig:causal}
\end{figure}
\begin{figure}
    \centering
    \includegraphics[width=\linewidth]{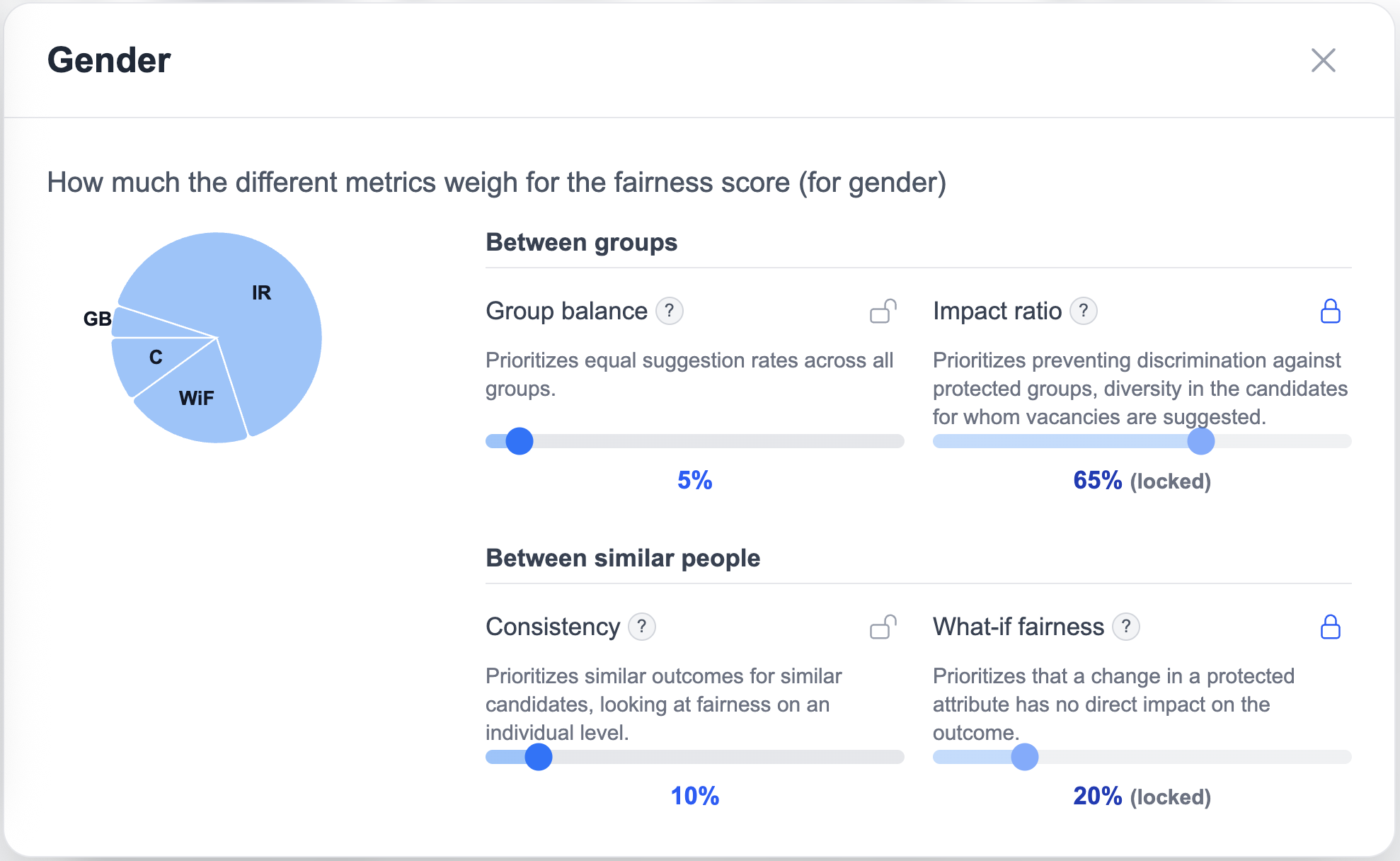}
    \caption{The weight component of the \textit{Experiment Page}. Metrics' impact on the overall fairness score can be adapted by changing the weight for that metric. This can be done using the sliders or presets (not visible here). This shows the component for gender, but it can be adapted either globally or for any specific attribute. }
    \label{fig:db_weights}
\end{figure}
\subsubsection{Data and recommendation algorithm.}
FairAware visualizes the fairness of a recommendation algorithm that suggests matching vacancies to job seekers. The underlying dataset is synthetic and was generated for this study (see Appendix A for details). We constructed 10000 candidate records with seven attributes selected together with our HR partner as the most relevant for fairness assessments in this domain: age, gender, education level, nationality, work experience, employment gap, and target-language fluency. Common biases (for age, gender, and language fluency) were incorporated based on interviews with HR employees, enabling us to verify the results. Vacancies were grouped by region and sector following the local public employment service's job-cluster taxonomy \cite{vdab_clusters}, as requested in user interviews. The recommendation algorithm matches candidates to vacancies on a skills similarity score and returns a binary ``suggest this vacancy/do not suggest" output per pair, which is what the four fairness metrics in FairAware are computed over. It is therefore a job-matching recommender that shows vacancies to candidates, not a résumé screener or a candidate ranker: bias manifests as unequal access to vacancy suggestions.

\subsubsection{The FairAware dashboard.}
FairAware (Figure \ref{fig:dashboard}) is the interface we built to present the fairness of the recommendation algorithm described above (or any other algorithm) to non-technical users. 
It was co-designed with HR domain experts through biweekly meetings with an industrial partner with over 15000 employees, semi-structured interviews with inclusion experts (n=4), a feedback session with the company's innovation board (n=6), and five semi-structured interviews with HR employees and a job seeker (n=5).
Transcripts were thematically coded following a hybrid inductive-deductive approach \cite{Braun2006}. From the resulting codebook (Appendix E), we identified three main themes, which we translated into the following design goals:
\begin{description}
    \item (DG1) Manage cognitive load, operationalized by letting users drill down from the overall score to metric-specific graphs, separating the \textit{Home}, \textit{Experiment}, and \textit{Causal Analysis Page}, and hover-cards with one-sentence metric explanations.
    \item (DG2) Protect candidate privacy by showing only aggregated group-level data and warning when a group is too small to show significant results.
    \item (DG3) Support stakeholder agency, as participants repeatedly asked to run their own queries in addition to the pre-selected results. This is implemented through user-defined attribute binning (e.g., customizable age ranges or employment gap thresholds), free attribute selection on the home tab, and global or per-attribute weight sliders (with optional preset profiles), to configure their own metric preferences. 
\end{description}

The dashboard presents biases according to four fairness metrics, two at the group level and two at the individual level, selected based on prior literature \cite{Taka2018, Lee2019_webuildai, Kearns2018, Verma2018, Luo2025_EARN, Waller2025}: For each metric, the participant-facing label is followed by the formal name in parentheses.
\begin{description}
    \item[Group balance] (statistical parity) compares the vacancy suggestion rate of a given group to the average vacancy suggestion rate. A value of zero indicates parity. Chosen for its alignment with human intuition \cite{Cheng2021_FEI}.
    \item[Impact ratio] (disparate impact) is the suggestion rate of a group divided by the rate for the most privileged group. Values below 0.8 are typically flagged as discrimination. Chosen for its familiarity in the organizational context.
    \item[Consistency] (consistency) measures, for each candidate, what fraction of their k-nearest neighbors in feature space received the same recommendation. The dashboard reports the mean for the group. When the user clicks on a result, it is complemented with Low-Individual-Consistency-Candidates (LICC, the absolute count of inconsistently treated people in the group \cite{Waller2025}) in the more detailed view.
    \item[What-if fairness] (counterfactual fairness) asks `if a candidate's sensitive attribute were flipped (e.g., female to male), would the recommendation change?'. The metric reports the proportion of candidates for whom the recommendation would flip.
\end{description}

There are five tabs, each with its own functionality available to participants:
\begin{itemize}
    \item The \textit{Home Page} (Figure \ref{fig:dashboard}): the main page of the dashboard, where users can see the fairness for the three different demographic groups that experience the highest bias, according to an ``overall fairness score'. Below it, users can select any attribute (e.g., age) to see fairness results for specific groups, or select an intersection of two attributes (e.g., `women aged 18-33'). For each group, the fairness is shown as the scores of the four fairness metrics, as well as an overall `fairness score'. This fairness score is computed by rescaling the four metrics to a shared scale that represents how `good' the value is for that metric, and then computing the weighted sum of these four values, based on the metric priorities that the user has chosen. For every metric value, hovering shows a one-sentence natural-language explanation, and clicking opens a detailed graph (e.g., for consistency, a bar chart showing what fraction of similar candidates received the same recommendation per group). 
    \item The \textit{Information Page} (Figure \ref{fig:infopage}): the start page upon first use, containing explanation cards for each of the four metrics, with a one-sentence definition, a short explanation, and an example. This tab does not change with user input.
    \item The \textit{Experiment Page}: lets the user 1) re-bin attributes (e.g., adapt the age ranges or the number of months that defines an employment gap) and 2) re-weight the four metrics according to their preferences (Figure \ref{fig:db_weights}), to configure their own fairness policy. Changing weights immediately updates the overall fairness score on the \textit{Home Page}. 
    \item The \textit{Causal Analysis Page}: contains a directed graph (Figure \ref{fig:causal}) over the seven attributes and the `fit score' (the similarity score of how well the candidates' skills match a vacancy, used to compute the binary match). Edge thickness encodes the strength of the relationship. Users have the option to give feedback on edges that they find unacceptable, but this does not change anything on the other pages. This page can be used when a user finds a bias and wants to know what other attributes might have influenced this (e.g., if we want to solve the work experience bias, should we also look at gender?).
    \item The \textit{Extra Data Insights Page}: contains data distribution histograms and a per-metric ranking of biased groups. This provides more information on the dataset.
\end{itemize}

We conducted a face-to-face pilot with five participants from an HR company (two recruiters, two inclusion-focused HR employees, one job seeker) prior to the study. This surfaced five design issues: confusion about the way that metrics were weighted, unnoticed small-group warnings, the recurring suggestion to separate exploration from the home page, a misinterpreted metric calculation from the example, and low visibility of the gradient bars that indicate a metric value's severity. This led to five changes in the final tool: a more explicit explanation of the weighted `overall fairness score', a split to separate the \textit{Home Page} and the \textit{Experiment Page}, traffic-light background colors added to the gradient bar, hidden results for small groups (only name, size, and `not enough data' label shown), and using unequal group sizes in the metric examples.

\subsection{Study Setup}
\label{subsec:setup}
\subsubsection{Participants.} 
We recruited 70 participants via Prolific, a platform to recruit online participants in studies \cite{prolific2026}. This sample included  35 HR professionals (24 women and 11 men, aged 21-57, M=33.6, SD=8.5) and 35 job seekers (11 women and 24 men, aged 18-58, M=26.6, SD=8.2). Both groups were included as they represent two sides of algorithmic hiring with different stakes.

Participants had to be fluent in English, with at least 80\% task approval ratings, and aged 18+. Prolific screeners were used, complemented with our own screener questions (Appendix B) to validate appropriate HR expertise.
Two participants were excluded and replaced for quality reasons. Participants were compensated at £9/hr.

\subsubsection{Procedure.}
\begin{figure*}
    \centering
    \includegraphics[width=\linewidth]{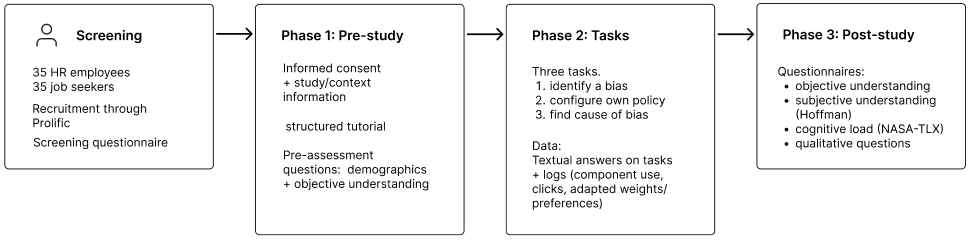}
    \caption{Study setup of the mixed-methods study}
    \label{fig:prolific_overview}
\end{figure*}

The study was conducted online and lasted around one hour (M=70.89 minutes). After the screening, it followed three phases (Figure \ref{fig:prolific_overview}). In Phase 1, participants gave informed consent, went through a structured dashboard tutorial (interactive pop-ups that they had to click through, M=8.92 min), and completed pre-study questionnaires (M=25.20 min) covering demographics, familiarity with fairness metrics, statistics, and interpreting graphs, as well as objective understanding (OU). In Phase 2 (M=16.61 min), they performed three tasks: 1) they identified the most biased group according to an attribute of choice, 2) they configured metric weights to reflect their own fairness policy, and 3) they investigated the cause behind a bias. Finally, in Phase 3, they completed a post-study questionnaire repeating the OU instrument and the NASA-TLX cognitive load questionnaire \cite{nasatlx}, a four-question subjective understanding scale adapted from \citeauthor{Hoffman2023}, and five open-ended feedback questions on what helped them the most/least to understand the content, sources of confusion, accessibility, and any further thoughts (M=16.61 min). The full questionnaires are available in the supplementary materials (Appendix C).

\subsubsection{Data collection and analysis.} 
We collected objective understanding, subjective understanding, cognitive load, qualitative responses, and logging data, and analyzed each as follows.

We measured \textit{Objective Understanding (OU) through mental model scores.} 
As no validated instruments exist for measuring mental models of fairness metrics, we constructed one for FairAware, inspired by previous research \cite{Lakkaraju2016, Bhattacharya2024, Bhattacharya2025, Kulesza2010, Kulesza2012, Kulesza2013, Kulesza2015, Nimmo2024, Jalali2023, Ribeiro2016, Cheng2019, Bove2022, Lim2009, Hase2020}. 
First, the nine items (12.5 points in total) are split across three question types, balanced to reflect increasing depth: fact (16\%), simulation (36\%), and articulation (48\%) questions. None of the question types accounts for more than 50\%. 
72\% of the score touches the weighted fairness score, as that is the central concept of FairAware, and 16\% explicitly asks about metric definitions, although understanding the metrics is a prerequisite to answering some of the other questions well. 
Second, multiple-choice questions had at least four options and an `I don't know' option to reduce guessing, and six out of nine questions were open-ended to allow for a more accurate estimation of respondents' mental model.
Third, item difficulty was calibrated using a pilot to avoid floor and ceiling effects.
Open-ended items were scored with rubrics: two researchers independently scored 10\% of responses (weighted Cohen's $\kappa=0.728$, substantial agreement \cite{cohens_kappa-landis_koch}) before one researcher coded the rest.
The same instrument was used pre- and post-task. Pre-task (median=39.60\%, SD=23.10) and post-task (median=43.2\%, SD=23.01) scores showed no ceiling or floor effects. We define \textit{Learning Gain} (LG) as the post-task OU score minus the pre-task OU score. OU and LG were tested with paired t-tests or Wilcoxon signed-rank tests when non-normal, and group differences with independent t-tests or Mann-Whitney U.

\textit{Subjective Understanding (SU)} was measured post-task on four 7-point Likert items adapted from Hoffman et al.'s \cite{Hoffman2018, Hoffman2023} scale, covering predictability, intelligibility, and utility. 

\textit{Cognitive load (CL)} was measured post-task using the six NASA-TLX items on a 7-point Likert scale. Correlations between CL, SU, familiarity, and OU were all measured using Spearman's $\rho$.

\textit{Qualitative responses} were coded thematically using a hybrid inductive-deductive thematic analysis \cite{Braun2006} by two researchers on 10\% of the data (validated employing unweighted $\kappa=0.871$) before the remaining coding was done by one researcher. The responses consisted of the five feedback questions and three open-text answers for the three tasks the participants completed. These task responses were filled in while they were performing the tasks. For the first task, they were asked to write down the attribute they found most important in an assessment, and the group that experienced the most bias according to that attribute. The second task asked them to configure their ideal fairness policy and then briefly describe how they approached this. This allowed us to see whether they had clear preferences or priorities, and, if they did not change much, whether that was due to being content with the current configuration or to not understanding how to adapt it. The third task asked them to investigate what other attributes might have influenced the bias identified in the first task. They had to write down one of those attributes and briefly explain how they knew this.

\textit{Logging} via Microsoft Clarity and our own logs captured time per study phase, hovers, clicks on metric results, tab visits, attribute searches, and parameter changes (the latter was needed to score task 1, since changed weights alter the fairness score).
Finally, page blurring and fast returns were tracked as a heuristic for genAI use. Logging results were tested against OU using Spearman's $\rho$, and between groups with t-tests or Mann-Whitney U. Throughout, statistical significance is denoted as \textdagger $p<.10$, * $p<.05$, **$p<.01$, and ***$p<.001$. 

\section{Results}
\label{sec:results}
\begin{figure*}
    \centering
    \includegraphics[width=\linewidth]{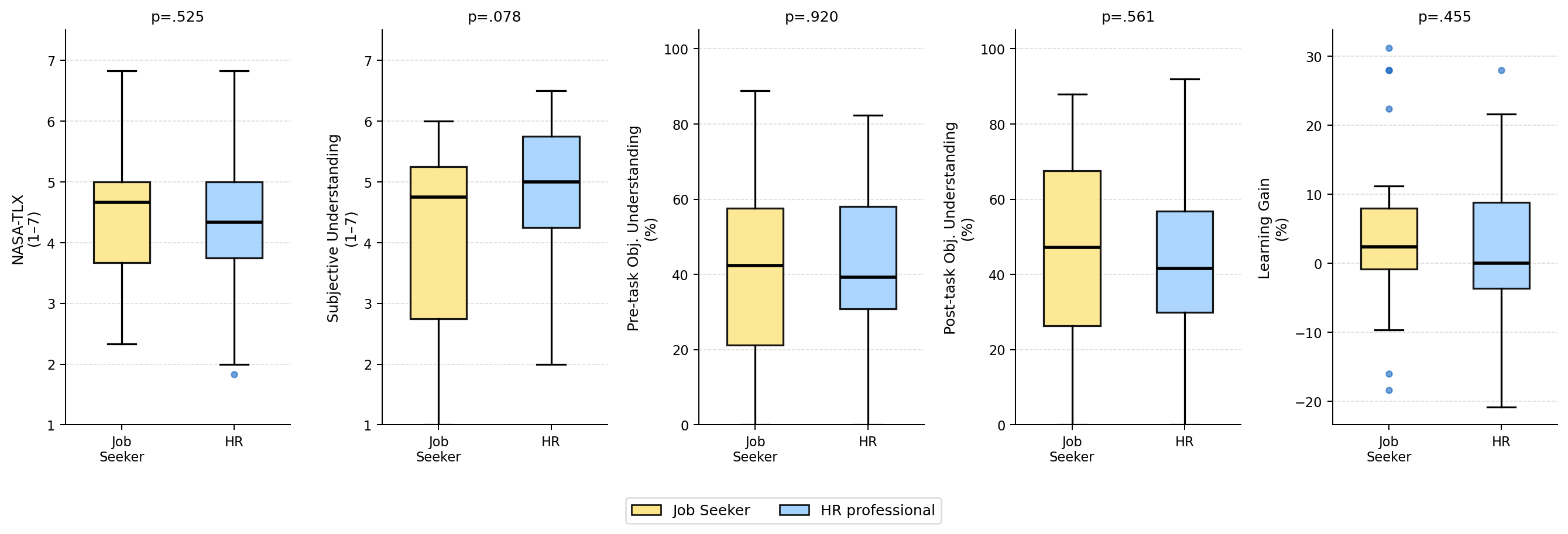}
    \caption{A comparison between job seekers and HR employees on cognitive load, subjective understanding, pre- and post-task objective understanding scores, and the learning gain, which is measured as the change in objective understanding score. None of the differences are significant, implying similar outcomes across stakeholder types.}
    \label{fig:js_vs_hr_boxplots}
\end{figure*}

We organize the results by research question (RQ). Across all measures and both RQs, job seekers and HR employees did not differ significantly (Figure \ref{fig:js_vs_hr_boxplots}); we therefore report the combined results below.

\subsection{RQ1: What Aspects Affected Users' Understanding and Ability to Find Biases?}
\label{subsec:results_rq1}
\subsubsection{Subjective understanding tracked confidence, not comprehension.}
\begin{figure*}
    \centering
    \includegraphics[width=0.6\linewidth]{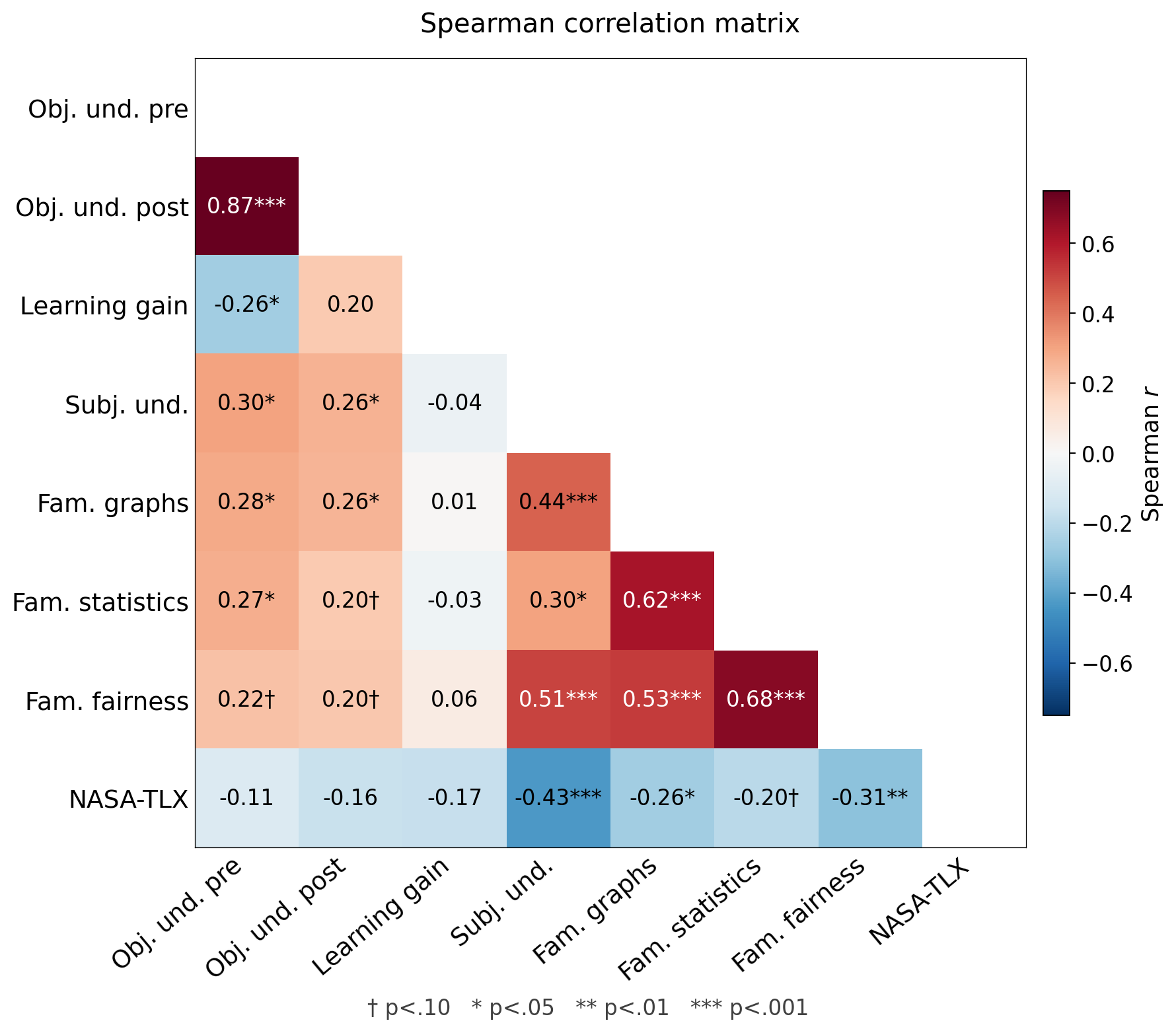}
    \caption{A heatmap of the correlations.}
    \label{fig:full_heatmap}
\end{figure*}

The strongest pattern in our data is a divergence between how well participants thought they understood the tool (SU) and how well they actually did (OU). SU correlated only weakly with post-task OU ($\rho=0.262, p=.028$*). It correlated strongly with prior familiarity with graphs ($\rho=+0.44, p<.001$***), statistics ($\rho=+0.30, p=.012$*), and fairness metrics ($\rho=+0.51, p<.001$***), and negatively with cognitive load ($\rho=-0.43, p<.001$***) (Figure \ref{fig:full_heatmap}). 
In other words, participants felt they understood the tool when it looked familiar, and felt they did not when they were overwhelmed, independently of whether they had actually understood the metrics and overall fairness score. The SU-OU correlation held for job seekers ($\rho=0.342, p=.044$*), but was weaker for HR employees ($\rho=0.295, p=.086$\textdagger).

Looking at the participants at the extremes of the SU-OU gap\footnote{Computed per participant as $(\frac{(SU-1)}{6}-\frac{(post-task OU)}{12.5})$, rescaling both to [0,1] before subtracting.} clarifies this pattern. On the high-SU/low-OU side, P11 and P20 scored 4/5 and 5/5 on the HR pre-screening but often gave superficial OU answers (e.g., answering `explain why you made these matches' with `because it is very clear'), showing overconfidence and lower engagement; 
P64 and P68 gave more elaborate answers, but misunderstood specific metrics, noting a low affinity with numbers and feeling overwhelmed by the amount of information as reasons for a lower OU score.
The reverse pattern, low-SU/high-OU, showed self-critical assessments rather than weak performance. P36 and P62 lost points due to hesitancy (P36: ``I did not made [sic] them because I am not sure how to"), despite otherwise insightful answers, and P43, P52, and P59 rated their SU moderately while providing good OU responses. They reported time pressure and information density as challenges, rather than confusion.

\subsubsection{Familiarity helped subjective understanding, but not objective understanding.}
Prior familiarity correlated only weakly or not at all with OU (graphs: $\rho=+0.26, p=.032$*; statistics: $\rho=+0.20, p=.099$\textdagger; fairness metrics $\rho=+0.20, p=.098$\textdagger).
Taken together with the SU findings above, this shows that familiarity improves SU but not OU. 
Cognitive Load shows the same pattern, not correlating with OU ($\rho=-0.17, p=.173$) despite the SU correlation.

These two patterns are also connected to a possible confound: that low OU reflects the effort of navigating an interface that presents four metrics and a weighted score at once, rather than the understanding of the metrics themselves. However, if navigation difficulty were the main factor, we would expect it to lower OU alongside SU and an increasing CL, yet CL did not correlate with OU and familiarity improved SU but not OU. This makes a pure navigation-difficulty explanation unlikely, though it does not entirely rule out its contribution.

\subsubsection{Most participants were able to spot a bias, but not because of understanding it.}
Despite the SU-OU divergence, 42 of 70 participants correctly identified the most disadvantaged group for an attribute of choice. Fourteen gave unclear answers, typically naming the attribute (`gender') instead of the group (`female'), and 14 were incorrect (vague answer beyond attribute-group confusion n=8, wrong group n=4, named a metric n=1, `I don't know' n=1). 
The biggest predictor of success was time on task ($p=.011$): neither pre- nor post-task OU, SU, nor CL predicted who correctly identified the bias.

As for participant characteristics, the 42 participants who identified the correct group differed from the 28 who did not in age (Mann-Whitney $U = 393.5, p = .020$), with successful participants being younger. However, there were no statistical differences due to gender ($\chi^2(1) = 0.54, p = .464$) or stakeholder group ($\chi^2(1) = 0.54, p = .464$). Within HR professionals, years of HR experience was unrelated to success ($\rho = -.152, p = .385$). Aside from a modest age effect, success in finding the bias depended on engagement with the tool, not on who the participant was.

This is consistent with the qualitative pattern that the tool has a learning curve, with early confusion (especially conflating groups with attributes during the first task) that improved with longer interaction. Bias identification after a first short session should therefore not be treated as a decision-ready result; appropriate use of the tool requires more interaction time.

Furthermore, read together with the SU-OU divergence, a non-technical participant can read the interface and point at the worst group, without forming an accurate mental model of why a group is treated worse or what the metrics measure.

Bias-spotting appears to be robust to understanding gaps for the same reason it is robust to expertise levels, as the tool shows which groups score worst at a glance, so the task is decoupled from understanding the underlying metric. A fairness assessment tool can thus be usable for high-level bias detection yet unsafe for higher-stakes participation, because the user might look confident and pick the right group at first sight, but their underlying understanding might be incorrect, meaning that more complex tasks could go wrong too. Any subsequent task that requires the user to trade off metrics, justify a choice, or deliberate with others might depend more on the objective understanding than the bias-spotting accuracy. 

\subsection{RQ2: How Do Users Employ the Dashboard to Advance Their Understanding and Find Biases?}
\label{subsec:results_rq2}
\subsubsection{Dashboard use produced a small gain in objective understanding.}
OU scores varied widely and were low overall (pre-task: median=39.60\%, SD=23.10; post-task median=43.20\%, SD=23.01). Learning Gain was significant across the full sample (Wilcoxon: $W=592.0, Z=-2.040, p=.041, r=0.244$*), a change driven by job seekers (Wilcoxon: $W=147.0, Z=-1.979, p=.048, r=0.335$*) but not HR employees (paired-t test: $t(34)=1.000, p=.325, d=0.169$).
The gain is modest in absolute terms but informative given the conditions: Phase 2 lasted 16.61 minutes on average, and the OU instrument included several deliberately challenging items. That a 15-minute first encounter produced a measurable increase in OU suggests that participants' OU would increase over time.

\subsubsection{Helpfulness reports showed what logs or dashboard use miss.}
Qualitative responses identify the \textit{Information Page} as the main page for understanding the content: 28 participants explicitly named it as one of the most helpful components, often noting the need to revisit explanations because of the high amount of new information (e.g., P47: `` I had to go back and look at the explanations of what each metric is a few times."). Naming it did not correspond to scoring better on OU (M=45.0\% against a rest-of-sample M=45.3\%). The \textit{Information Page} is the page participants default to, but using it does not separate participants who built a better mental model of the metrics from those who did not.

The \textit{Home Page}, named by 22 participants as most helpful, did separate participants on OU: those who named it as most helpful scored on average M=58.1\% against M=39.3\% for the rest of the sample (two-sided Welch's: p=0.002**, d=0.86, Holm-adjusted p=0.008**). Their open responses described their understanding of the overall fairness score and how the different metrics contributed to it. As the \textit{Home Page} summarizes most of the content, participants had to grasp it to find it helpful.

The \textit{Experiment Page} was the most polarizing: 14 participants named it as most helpful, generally as the place where the weighting mechanism started making sense (e.g., P34: ``the experiment tab helped me see how changing the weights of the metrics affects the fairness score, [...] easier to understand in a practical way."
), and 11 named it as least helpful, generally because they had not yet had time to digest the metrics and the \textit{Home Page} before being asked to manipulate it (e.g., P36: ``The experiment. [...] I had too little experience.").
An exploratory analysis shows that polarization is not the same across stakeholder groups: 11 of 35 job seekers named the \textit{Experiment Page} as most helpful, against only 3 of 35 HR employees, even though the groups did not differ on OU, SU, or cognitive load. We interpret this as a difference in what the two groups found informative rather than an indication of how much they understood. HR participants might be more used to reading an overview report, whereas a job seeker might put more emphasis on figuring out what policy they should prefer. 
The response that indicated that a participant did not feel ready to change the weights suggests that careful sequencing of the task is important. Interactive configuration can be valuable, but only once the static overview is fully internalized.

22 of 70 participants reported some form of overwhelm, substantially more often from job seekers (16/35) than from HR employees (2/35), which we take as another way the experience of the dashboard differs between the groups despite their similar OU, SU, and CL scores. 

\subsubsection{Self-reported usefulness did not directly translate to actual usage.}
The interactions with the \textit{Causal Analysis Page} showed some interesting contradictions. Sixteen participants named it as the least helpful, and only ten (five from each group) used it for the task that asked them to identify what other attributes might have influenced a bias, the task for which it was built. Moreover, of the seven participants who named it `most helpful', only two actually used it for that task, and two others could not correctly interpret the graph on the OU questionnaire. 
This exemplifies the SU-OU gap identified above, now at the component level: participants who could not interpret the graph nevertheless believed that it helped them, and five participants who reported that it was hard to interpret the graph interpreted it correctly (scoring 2/2 on OU). 

\subsubsection{Exploration depth varied, repeating explanations remains valuable either way.}
Usage logs did not correlate with OU, SU, or cognitive load. The qualitative responses help explain why.
First, exploration depth varied widely. At one end, some participants moved quickly through the tutorial without actively engaging the hover and click features, and when uncertain, they returned to the general explanations in the \textit{Information Page} instead of reading the in-context details. 
Others fully explored the tool, including the \textit{Causal Analysis Page} and the \textit{Extra Data Insights Page}. 
Hover and click counts did not serve as strong predictors here, because they conflate different participant strategies: curiously exploring details or being confused and seeking explanations by hovering result in the same behavior. This makes it difficult to distinguish between different ways participants use the available explanations. 
Second, the participants who moved through the tutorial quickly did not do so because they already understood the tool; the SU-OU pattern shows that several were overconfident. As a result, participants valued the explanations remaining accessible, as they reported in the feedback questions. A takeaway for tool design is that tooltips, tutorials, and hover-explanations should remain visible throughout the usage, not just at the start (e.g., as participants noted: `I was confused at first as to what the what-if was comparing until I hovered over it.', `more accessible: Having access to the tutorial').
 
\section{Discussion}
\subsection{Implications} 
\subsubsection{Beware of and counteract Mount Stupid.} 
We found that Subjective Understanding rose with familiarity and decreased with Cognitive Load, while Objective Understanding did neither. This aligns with the miscalibration effects documented in the broader psychology literature \cite{Krugeretal1999}, where people who feel they understand a domain are not necessarily the people who actually do. This raises a design concern for fairness assessment tools: an intuitive-looking or overly simplified interface might lower cognitive load and difficulty of using it, but at the cost of reducing participants' effort to verify or improve their understanding. Objective Understanding cannot be ensured by simplified UI design; it needs to be checked explicitly as well. 

Measuring and reporting both constructs is thus required. Familiarity with fairness metrics was the strongest correlation with SU and was unrelated to OU, so the users most at risk of unwarranted confidence might be the 'experts' that end up using the tool. Verification of the level of OU is thus important in any setting. This could be achieved through lightweight, quiz-like prompts embedded in the workflow, similar to those in our mental model questionnaire. In practice, these could be short, auto-scorable checks (e.g., a pop-up with ``If this group's balance worsened, the fairness score would go: up/down/stay the same/I do not know"). Responses could then be used to mitigate areas where understanding is compromised, perhaps through surfacing short explanations. This could identify flawed mental models early on and counteract them. Personalized adjustments could also be made to the interface based on the user's OU, mirroring work in personalized XAI \cite{conati_toward_2021}.

\subsubsection{Dashboard use does not ensure users understand fairness metrics.}
Hover counts, click counts, and page visits did not predict Objective Understanding, possibly because they do not communicate the user's goals:  the same hover can mean ``I am lost" or ``I am curious". This means that help and explanation features should remain visible and accessible from multiple places to provide multiple educational opportunities when they need it. Second, perceived component helpfulness was partially decoupled from actual usage. Of the seven participants who reported the causal graph as most helpful, only two actually used it for the bias-cause task. Similarly, reporting the \textit{Home Page} as most helpful correlated with higher OU. More work is needed to identify truly helpful pages and explanations for understanding fairness metrics.

\subsubsection{Algorithmic bias finding is achievable but takes time.}
Non-technical stakeholders can perform basic fairness assessments with an appropriately designed tool, but the only significant predictor of success was task duration, not OU or SU. However, increasing the time would also likely increase cognitive load, so it is a delicate balance to be struck. 
 
This might necessitate that organizations using fairness assessment tools need to conduct shorter, repeated assessments over time, rather than a longer single session. This would also allow for developing clearer fairness preferences through interaction and learning, and for evolving them, as they are not static \cite{AlRossais2025Time-Evolving}. 

\subsubsection{Consensus-building frameworks need to account for miscalibrated participants.} 
Frameworks like WeBuildAI \cite{Lee2019_webuildai} assume that stakeholders bring individual understanding into group deliberation and that stronger opinions might reflect deeper engagement. However, our results complicate that assumption, because some stakeholders may hold strong opinions based on an incorrect mental model and therefore low Objective Understanding. For example, some people might believe that a high impact ratio means the algorithm is fair when it actually indicates the opposite. This could steer a group discussion towards incorrect conclusions. We argue that consensus-building frameworks thus need to be designed for \textit{calibrated participation}, for example, by gating opinion weighting with a brief understanding check.

\subsubsection{Transferring FairAware to other domains.} Although FairAware's content is domain-specific, its structure is generalizable: the four metrics, the rescaled weighted score, the drill-down, and the causal view are defined over any binary decision with a sensitive attribute. The tool could therefore be readily translated to other settings based on tabular data, e.g., loan approval or benefits triage, by replacing the attribute set and the group levels. However, the metrics and group levels have to be based on co-design, since which groups and metrics are meaningful is something that depends on domain context \cite{Cheng2019, Lee2019_webuildai}. 

\subsection{Future Work}
Three potential future work directions follow from our results. First, longitudinal, real-world studies are needed to see how mental models of non-technical stakeholders evolve with repeated use of fairness assessment tools, whether bias identification accuracy increases, and whether the SU-OU gap decreases or not.
Second, our finding that bias-finding accuracy is based on time rather than understanding suggests further research into the effects of more complex tasks, such as policy configuration, metric trade-offs, or justification of assessment decisions on SU, OU, and the ability to assess biases.  
Third, the consensus-building implication motivates empirical research that investigates how understanding affects deliberations of fairness decisions within groups.

\subsection{Limitations}
Four shortcomings qualify these results. First, some participants completed the tasks quickly, which might limit engagement with the complex concepts we presented in the study. Second, we presented only four fairness metrics in our dashboard. Although these were selected during co-design as the most relevant, this work could be extended to include more fairness notions, some of which could be user-specified \cite{Luoetal2026_ithinkthisisfair}. Third, the study was conducted online via Prolific, in a one-time session with no consequences attached to participants' judgments, using synthetic data. While we used screening to ensure the quality of responses, an online setting offers less control than an in-person study. Indeed, authentic data might be more difficult to assess but provide a high-stakes setting, which could influence engagement and calibration. More work is warranted to study the use of tools in practice. Fourth, no validated questionnaire exists for measuring mental models of fairness metrics or assessment tools. Our instrument discriminated well between levels of understanding without ceiling or floor effects, but absolute scores cannot be directly compared to those from other mental-model questionnaires. Standardization of OU measurements is desperately needed.

\section{Conclusion}
We tested a common assumption in algorithmic fairness work: that the non-technical stakeholders asked to choose between or interpret fairness metrics actually understand those metrics. To do this, we (1) co-designed FairAware, a fairness-assessment dashboard for non-technical stakeholders in the HR domain, (2) built a mental-model instrument grounded in prior literature to measure their objective understanding, and (3) ran a mixed-methods study with 35 HR employees and 35 job seekers.

The assumption does not always hold. Across 70 participants, subjective understanding correlated with prior familiarity but only weakly with objective understanding; HR employees were not better calibrated than job seekers; and the only predictor of correctly identifying a biased group was time spent, rather than understanding.
Three implications follow: 
\begin{itemize}
    \item For tool developers: embed lightweight comprehension checks into the workflow rather than relying on user-reported confidence. These checks can then be used to assess the reliability of these users' takeaways from the tool, as well as serve to provide short, targeted educational opportunities where miscomprehension occurs frequently.
    \item For participatory fairness research: when aggregating preferences into a shared policy, weight stakeholder input by calibrated comprehension rather than familiarity or subjective understanding. When conducting group discussions, take the objective understanding into account as well (e.g., by first testing it and providing a short education on miscalibrated aspects).
    \item For deployment: treat single-session assessments as preliminary. The EU AI Act's `meaningful human oversight' requirement is not met by a confident user who has quickly read the interface for the first time, but not grasped the metrics.
\end{itemize}

Our work contributes a stakeholder-facing tool and empirical evidence that comprehension checks are needed before stakeholder input can be trusted. Our mental model instrument offers a starting point that others can adapt to their own tools and metrics.
Together, these contributions support the more responsible integration of stakeholders into fairness assessments and, through that, more responsible AI.

% If so, you can uncomment the following line and adjust the path to include it.

%\section{Ethical statement}

\section{Acknowledgments}
% Please limit to no more than 3 sentences
This work has been funded and supported by the imec.icon CAPTURE project and the Flanders AI Research Program (FAIR), with project support from imec and VLAIO (Flanders Innovation \& Entrepreneurship)(Grant No. HBC.2024.0220). Furthermore, we would like to thank all participants for their time and valuable feedback.

\bibliography{references}
\newpage\appendix
\section{Supplementary Materials for `Spotting (and Missing) Algorithmic Bias: Investigating User Understanding in a Fairness Assessment Tool'}
\subsection{Appendix A: Dataset Construction}
\label{app:dataset}
We constructed a synthetic dataset of 10.000 job candidates and 34 vacancies, representative of the local labor market. Each candidate is characterized by age (uniform, 18–67), gender (male/female/other; 48/47/5\%), nationality (12 categories with local-majority prevalence), education level, years of experience, resume gap, and a skill set drawn from a domain-specific pool of 18 vocational skills.  
Candidate–vacancy fit is measured as skill overlap: score = $\frac{|candidate skills \cap required skills|}{|required skills|}$, yielding a continuous score in [0, 1].

\textbf{Nationality and language} Nationality is drawn from 12 categories via a single weighted categorical draw, independent of every other attribute (no correlation with age, gender, or region): Belgian (40\%), Dutch (30\%), German (10\%), French (5\%), Spanish (5\%), Italian (2\%), Turkish (2\%), Polish (2\%), Indian (1\%), Syrian (1\%), Ukrainian (1\%), Other (1\%). These nationalities are chosen to represent a plausible composition of the regional labor market. Spoken languages are then generated conditional on nationality: Belgian candidates receive Dutch, English, and French as base languages; Dutch candidates receive Dutch and English; candidates of all other nationalities receive their nationality's language as a base. Each candidate additionally receives 0–3 languages sampled from a weighted 12-language pool (Dutch, French, English, Spanish, German, Italian, Turkish, Polish, Indian, Syrian, Ukrainian, Other), with English, Dutch, and French weighted highest.

\textbf{Education, experience, and vacancies}. Education level ({none, high school, bachelor, master}) is drawn conditional on age band and on whether the candidate speaks the local language (Dutch), as is described in more detail in the bias description below. Experience is sampled uniformly on [0, age-18], reduced by 5 years for bachelor's/master's holders with $\geq$5 years, to approximate time spent in education. Each of the 34 vacancies requires 10–12 skills sampled without replacement from the same 18-skill pool.

Group-level disparities are planted by construction. To simulate the real-life disparities that recurred in semi-structured interviews with HR employees, female candidates are assigned 2 fewer skills on average and have a 15\% higher probability of a resume gap. Candidates aged 54+ receive 2 fewer skills and a capped experience bonus, reducing their scores despite their long work experience. Candidates who do not speak the local language have slightly lower probabilities of higher education levels, as they might have had fewer opportunities for this. Because the direction and magnitude of each bias are known, the dataset enables ground-truth evaluation of fairness detection: a fairness assessment tool should flag the protected attributes for which disparities were introduced, and should not flag unbiased attributes as discriminatory.

\subsection{Appendix B: Participant Screening}
\label{app:screening}
\subsubsection{On the screening criteria}
The participant screening criteria table can be found in table \ref{tab:screening}. English fluency was deemed necessary because FairAware contains many textual explanations. If a participant cannot read the explanations, they will not be able to understand the content well. 
The 80-100\% approval rate was chosen in order to exclude participants who often turn in low-quality answers. For the HR group, the Prolific pre-screener `Employment role: Human resources' was used. This provides a first screening, but in order to have more control over the data quality, an additional screening was added. 
For the job-seeking group, multiple of Prolific's pre-screeners were added. The pre-screener `Unemployed (and job seeking)' was used, as the target group is people who are looking for a job. However, additionally adding the screener `Actively looking for a job' brought the participant pool down from 36.279 to 9.364. As this difference was high, the additional screener was added as well, to increase the chances of an accurate participant pool. Lastly, they had to pass a simple screener by answering 'Are you currently employed?'. This was added to avoid having employed participants due to an outdated Prolific status.
If the participant answered `yes' on this question, they were asked to withdraw from the study; if `no,' they could continue.
\begin{table}[ht]
    \centering
    \renewcommand{\arraystretch}{1.5} % 
    \begin{tabular}{l p{5cm}} 
        \toprule
        \textbf{Group} & \textbf{Prolific screeners \& own quality controls} \\
        \midrule
        \textbf{HR Group} & \small Fluent English; \newline 80--100\% Approval Rate; \newline Employment role: Human resources. \\        
        \midrule
        \textbf{Job-Seeking} & \small Fluent English; \newline 80--100\% Approval Rate; \newline Unemployed (and job seeking);\newline Actively looking for a job; \newline Excluded HR participant group. \\
        \midrule
        \textbf{Validation} & \small  Verification of job status/description in both an early multiple-choice question and open questions in the demographic questionnaire;\newline Attention checks and golden questions;\newline For HR group: 3/5 score on domain questionnaire.\\
        \bottomrule
    \end{tabular}
    \caption{Participant screening criteria}
    \label{tab:screening}
\end{table}

\subsubsection{Domain screening questionnaire}
\label{sec:hr_screening_questionnaire}
The questionnaire we used for screening the HR employees was combined from different sources.
The first question asks about their job role: if a participant's job is not directly related to hiring decisions or inclusion, they are redirected to Prolific. This was chosen to avoid the inclusion of too many, e.g., programmers in an HR company, as we want to prioritize employees in a job closely related to the potential use of such a recommendation algorithm.
The other five questions are related to their domain knowledge, to filter out participants who might not be fully honest on the first question and do not actually work in the domain.
The participants had to answer at least three of the five questions correctly to continue the study. This meant that there was only a 10,35\% chance of guessing correctly, which was deemed low enough to proceed. The reason we used a threshold is that the human resources domain is broad, and knowledge might differ slightly across job roles and countries. Expecting employees to answer all five correctly is thus deemed too strict.
The second question was proposed by a domain expert, whereas the third through the sixth questions were copied from a textbook on human resources in Canada \cite{Hapke2024Chapter}. The selection of questions was made based on difficulty and geographic scope (e.g., any questions that were specific to Canadian law were excluded).
The correct answers are shown in bold. If a participant answers C on the first question, they are immediately redirected to a page that leads them back to Prolific and asks them to withdraw.\\
\newline
\textbf{Q1. Which of the following best describes your involvement in hiring candidates? }
\begin{enumerate}[label=\Alph*.]
    \item I have a direct involvement in hiring decisions 
    \item I have a more high-level involvement with recruiting: DEI manager, member of an ethics or innovation board, or another inclusion-focused job in an HR company 
    \item I have no direct involvement in hiring decisions or ethics/inclusion in an HR context. 
\end{enumerate}
\bigskip      
\textbf{Q2. Why is it important to provide a clear explanation when recommending a candidate for a role?}
\begin{enumerate}[label=\Alph*.]
    \item To speed up the onboarding process once an offer is accepted 
    \item \textbf{It increases transparency and helps avoid potential bias or legal risks in hiring decisions} 
    \item It helps make the job advertisement look more attractive for future roles 
    \item It is mostly optional unless the hiring manager explicitly asks for it 
\end{enumerate}
\bigskip
\textbf{Q3: What is a key benefit of conducting a trend analysis in HR planning? }
\begin{enumerate}[label=\Alph*.]
    \item It helps in designing employee benefits packages 
    \item \textbf{It predicts future staffing needs based on historical data}
    \item It improves employee morale 
    \item It increases the number of job applicants 
\end{enumerate}
\bigskip
\textbf{Q4: Why is effective employee selection important for an organization? }
\begin{enumerate}[label=\Alph*.]
    \item It reduces the need for training 
    \item It ensures compliance with labor laws 
    \item \textbf{It leads to higher productivity, better performance, and lower turnover} 
    \item It speeds up the hiring process 
\end{enumerate}
\bigskip
\textbf{Q5: Which of the following is a key purpose of an aptitude test in the selection process?} 
\begin{enumerate}[label=\Alph*.]
    \item To assess a candidate’s honesty and integrity 
    \item \textbf{To evaluate a candidate’s natural ability to learn new skills}
    \item To measure a candidate’s knowledge about a specific job 
    \item To determine a candidate’s emotional intelligence 
\end{enumerate}
\bigskip
\textbf{Q6: What is an applicant tracking system (ATS) primarily used for?} 
\begin{enumerate}[label=\Alph*.]
    \item Conducting  face-to-face interviews 
    \item Providing job training to new employees 
    \item Creating job postings on social media 
    \item \textbf{Automating the initial review of job applications }
\end{enumerate}

\subsection{Appendix C: Questionnaires}
\label{app:questionnaires}
This appendix contains all the questionnaires used in the study. 
Pre- and post-task mental model questionnaires were the same (apart from attention checks or golden questions such as `please select 5 on this question' or `What is not related to this study? (a) fairness metrics (b) pineapples (c) weighted scores (d) biases (e) I don't know). Attention checks were used both during the tutorial and in the pre- and post-task questionnaires.

\subsubsection{Demographics questionnaire}
\textbf{1.} What gender do you identify with?
\begin{itemize}
    \item Woman
    \item Man
    \item Other
    \item Prefer not to say
\end{itemize}
\textbf{2.} What is your age? \\
\textbf{3.} Do you currently have a job? 
\begin{itemize}
    \item yes 
    \item no ($\rightarrow$ get directed to question 5)
\end{itemize}
\textbf{4.} Do you have any previous work experience related to human resources and/or ethics (fairness, inclusivity, ...)? Yes/no\\
What experience and how long?\\ ($\rightarrow$ get directed to end of demographics questionnaire)\\
\textbf{5.} What is your current job? How is that job related to human resources?\\
\textbf{6.} How many years of experience do you have in that job?\\
\textbf{7.} Do you have any additional experience in HR and/or ethics? If so, what experience?\\

\subsubsection{Familiarity questionnaire}
How comfortable are you with...\\
fairness metrics?\\
statistics?\\
interpreting graphs? (this included an example figure of a directed graph with three nodes)

\subsubsection{The mental model questionnaire, including scoring}
\begin{table}[h!]
    \centering
    \begin{tabular}{c|l|c}
        Question & Type & max score  \\
        \hline
        1 & fact & 1\\
        2 & fact & 1\\
        3 & articulation & 2\\
        4.1 & simulation & 2\\
        4.2 & articulation & 2\\
        5 & simulation & 1 \\
        6.1 & simulation & 1 \\
        6.2 & simulation & 0.5 \\
        7 & articulation & 2 \\
    \end{tabular}
    \caption{Question overview. The total score is 12.5 and is later converted to a percentage.}
    \label{tab:q_overview}
\end{table}
\begin{table}[h!]
    \centering
    \begin{tabular}{l|l|l}
        Type & score percentage \% & questions  \\
        \hline
        Fact & \textcolor{gray}{2/12.5 $\rightarrow$} 16\% & 1, 2 \\
        Simulation & \textcolor{gray}{4.5/12.5 $\rightarrow$} 36\% & 4.1, 5, 6.1, 6.2\\
        Articulation & \textcolor{gray}{6/12.5 $\rightarrow$} 48\% & 3, 4.2, 7 \\
    \end{tabular}
    \caption{Percentages of question types in the mental model score.}
    \label{tab:q_perc_types}
\end{table}
% --------------------  QUESTIONS -------------------
% Is the metric clear?
\textbf{1.} An impact ratio of 0.75 means that the unprivileged group... 
\begin{enumerate}[label=\alph*) , leftmargin=2cm]
    \item gets the relevant vacancy recommended (or succeeds) \textbf{at 75\% the rate of the privileged group} (e.g. 10\% of men get the suggestion, compared to only 7.5\% of women).
    \item gets the relevant vacancy recommended \textbf{at 25\% the rate of the privileged group} (e.g. 10\% of men get the suggestion, compared to only 2.5\% of women).
    \item has a \textbf{75\% chance} of getting a relevant vacancy recommended, \textbf{regardless of} the suggestions that \textbf{the privileged group} get (e.g. 75\% of women get this job suggestion).
    \item gets a relevant vacancy \textbf{recommended 1 out of 4 times} (e.g., 1 out of 4 women get this job suggestion).
    \item I don't know
\end{enumerate}

\textbf{2.} Explain in your own words what \textbf{consistency} and \textbf{impact ratio} mean (in the context of fairness metrics).\\
\textit{Rubric}\\
For both consistency and impact ratio:
\begin{itemize}
    \item +0.5 mark: explanation mentions that these are (fairness) metrics.
    \item +0.5 mark: explanation mentions that consistency focuses on unfairness towards individuals. 
    \item +0.5 mark: explanation mentions that the impact ratio focuses on unfairness towards groups. 
    \item +0.5 mark: explanation includes that consistency makes sure that similar people are treated similarly.
    \item +0.5 mark: explanation includes that the impact ratio compares a group to the most privileged group.
    \item 0 mark: incorrect or blank answer 
\end{itemize}
Total score = score/2.5\\

\begin{figure*}
    \centering
    \includegraphics[width=\linewidth]{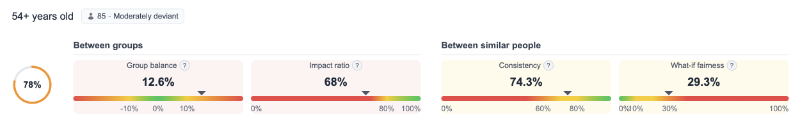}
    \caption{Weighted overview accompanying question 3.}
    \label{fig:q_gewogen_overzicht_full}
\end{figure*}

% Is the overall score clear?
\textbf{3.} Look at this screenshot of a weighted overview (figure \ref{fig:q_gewogen_overzicht_full}). Explain in your own words what the 78\% on the left means and how this number is calculated (high-level, just the concept, you don't have to do actual calculations).\\
\textit{Rubric}
\begin{itemize}
    \item +0.5 mark: explanation mentions that it is an indication of overall bias/fairness
    \item +0.5 mark: explanation mentions that higher is better
    \item +0.5 mark: explanation mentions it includes the four metrics' values for that group
    \item +0.5 mark: explanation mentions weights of the metrics
    \item -0.5 mark: incorrect information (capped at zero)
    \item 0 marks: incorrect or blank answer \\
\end{itemize}

\begin{figure*}[h!]
    \centering
    \includegraphics[width=0.8\linewidth]{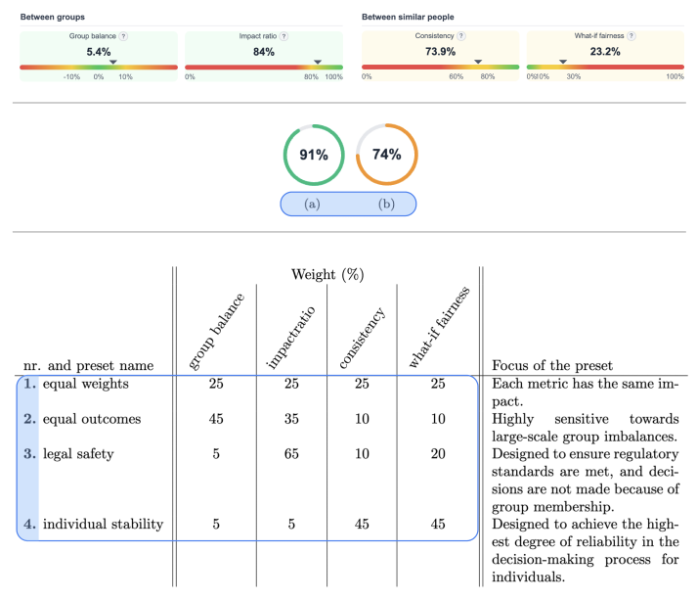}
    \caption{Accompanying figure for questions 4.1.}
    \label{fig:q_weighted_overview}
\end{figure*}

% weighted overview score, weights
\textbf{4.1} In the dashboard, there are some 'presets' available that give previously established weights to each metric.
In the image (figure \ref{fig:q_weighted_overview}), you see results for the four metrics. For these results, you see the \textbf{weighted fairness scores (a and b)} that belong to two of the \textbf{presets (1-4). Match the fairness scores a }(91\%)\textbf{ and b} (74\%)\textbf{ with their preset}, e.g. 'a3-b1'. 

\textbf{4.2} \textbf{Explain} why you made these matches (previous question): \\
\textit{Rubric}
\begin{itemize}
    \item +1 mark: explanation shows understanding that it is based on the metric priorities of each preset
    \item +0.5 mark: explanation for a mentions high weights on metrics with `green' results or low bias 
    \item +0.5 mark: explanation for b mentions high weights on metrics with `bad' results
    \item -0.5 mark: incorrect information on the weighting mechanism
    \item 0 marks: incorrect or blank answer \\
\end{itemize}
How confident do you feel in your answer? (7-point Likert scale) \\
\newline

\textbf{5.} The current group balance is $1.3\%$ and the weighted fairness score is 77\%. \textbf{What happens to the weighted fairness score if the group balance becomes -20\% instead of 1.3\%?} (you may assume that the results of other metrics remain the same and that they have equal weights)  
\begin{enumerate}[label=\alph*) , leftmargin=2cm]
    \item The weighted score will become \textbf{negative}
    \item The weighted score will \textbf{decrease} (e.g. to 70\%)
    \item The weighted score will \textbf{remain the same} (around 77\%)
    \item The weighted score will \textbf{increase} (e.g. to 85\%)
    \item I don't know \\
\end{enumerate}

\begin{figure}[h!]
    \centering
    \includegraphics[width=\linewidth]{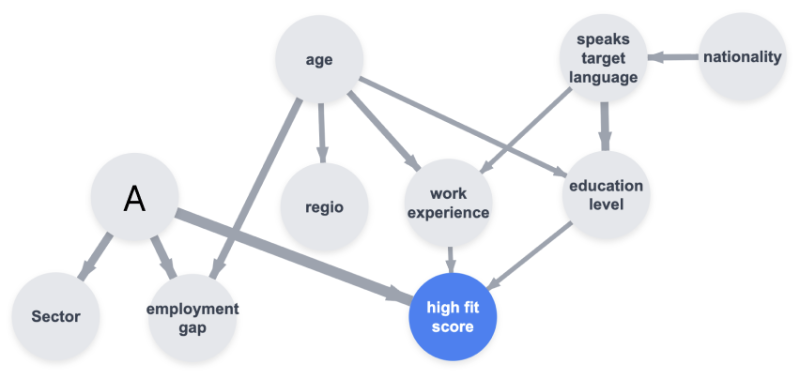}
    \caption{Causal graph accompanying question 6.}
    \label{fig:q_scg}
\end{figure}

% Did they look at the causal graph? Do they understand how it works?
\textbf{6.1} On the causal graph in this figure (figure \ref{fig:q_scg}): if attribute 'A' would change for a person, which attributes would be affected by that?
\begin{itemize}
    \item sector
    \item employment gap
    \item work experience
    \item speaks target language
    \item education level
    \item nationality
    \item region
    \item age
    \item high fit score
    \item none of the above
    \item all of the above
    \item I don't know\\
\end{itemize}
\textit{Scoring} 1 mark if all correct, 0 marks if not. \\

\textbf{6.2} Which of the attributes in the previous question would experience the \textbf{biggest influence}? 
\begin{itemize}
    \item attribute: \_\_\_\_\_\_\_\_\_\_\_\_\_\_\_\_
    \item I don't know\\
\end{itemize}
  
\textbf{7.} This dashboard uses multiple metrics (group balance, consistency, ...). Does that have an added value? Or can the same information be shown with one metric? Why do you think so? \\
\textit{Rubric}
\begin{itemize}
    \item +1 mark: Mentions that you cannot optimize for all metrics at the same time, or that one overarching metric cannot capture the same.
    \item +1 mark: Mentions that these metrics focus on different aspects. 
\end{itemize}

\subsubsection{Notes on the mental model questionnaire}
\paragraph{Question format} The mental model score is affected by the choice of question format (open-ended or multiple-choice). To validate this statement, the first two questions in the OU questionnaire ask whether the participant understands what the metric `impact ratio' means. The first question is in multiple-choice format, and the second is an open question. As the participant can still see the first question and its potential answers while answering the second question, we expected the participants who answered the first question correctly to score at least 0.5 on the second question, since they could obtain 0.5 marks for knowing it was a group metric, which was present in all the answer options. 
Interestingly, $57\%$ chose the correct multiple-choice answer pre-task, rising to $61\%$ post-task (table \ref{tab:ir-recognition-articulation}), but of those who chose the correct multiple-choice answer, $40\%$ scored 0 on the open-ended version (pre-task, table \ref{tab:ir-gap}). This means they picked the right multiple-choice option but could not demonstrate this knowledge when asked to provide the definition themselves, not even by copying the answer. Only $45\%$ of correct multiple-choice respondents received the full mark. This shows that the question format used to measure a mental model can have a large influence on the outcome. As expected, recognizing an answer yields higher scores than having to articulate it.

\paragraph{`I don't know' contradictions}
If we look at the same two questions in the pre-task OU questionnaire, four participants explicitly responded that they do not know what the impact ratio is. They did not provide an additional guess to its meaning either. However, of those four participants, only one had answered `I don't know' on the previous question (multiple-choice question about the meaning of impact ratio). In the post-task OU questionnaire, the same situation occurred: two participants answered that they did not know the meaning in the open-ended question, but none of them answered `I don't know' on the multiple-choice version.
That does not mean that no one selected `I don't know'. In the pre-task questionnaire, three participants selected `I don't know' but then guessed the open-ended question. Of these three, two answered the open-ended question correctly.
In the post-task questionnaire, four participants selected `I don't know' (only one participant overlapping with the pre-task questionnaire `I don't know' respondents). One obtained 0.5/1 on the open-ended version, the other three 0/1, indicating a correct self-assessment here.

\begin{table*}[h]
\centering
\small
\begin{tabular}{@{}lrrrr@{}}
\toprule
 & \multicolumn{2}{c}{Pre-task} & \multicolumn{2}{c}{Post-task} \\
\cmidrule(lr){2-3} \cmidrule(lr){4-5}
 & $n$ & \% & $n$ & \% \\
\midrule
\multicolumn{5}{@{}l}{\textit{MCQ (recognition)}} \\
\quad Correct          & 40 & 57.1 & 43 & 61.4 \\
\quad Incorrect        & 26 & 37.1 & 23 & 32.9 \\
\quad ``I don't know'' &  4 &  5.7 &  4 &  5.7 \\
\midrule
\multicolumn{5}{@{}l}{\textit{Open-ended (articulation), IR portion}} \\
\quad Full understanding (1.0)    & 29 & 41.4 & 28 & 40.0 \\
\quad Partial understanding (0.5) &  8 & 11.4 & 12 & 17.1 \\
\quad No understanding (0.0)      & 33 & 47.1 & 30 & 42.9 \\
\quad Mean score (SD) & \multicolumn{2}{c}{0.47 (0.47)} & \multicolumn{2}{c}{0.49 (0.46)} \\
\bottomrule
\end{tabular}
\caption{Impact ratio understanding: recognition (MCQ) vs.\ articulation (open-ended). The MCQ and open-ended questions are both fact questions that assess understanding of the impact ratio metric in a different format; the MCQ requires selecting the correct interpretation, while the open-ended question requires explaining it in one's own words.
Pre vs.\ post open-ended scores did not differ significantly (Wilcoxon $W = 98.5$, $p = .802$).}
\label{tab:ir-recognition-articulation}
\end{table*}

\begin{table*}[h]
\centering
\small
\begin{tabular}{@{}lrrrrrr@{}}
\toprule
 & \multicolumn{3}{c}{Pre-task (N=70)} & \multicolumn{3}{c}{Post-task (N=70)} \\
\cmidrule(lr){2-4} \cmidrule(lr){5-7}
Open-ended score & \multicolumn{1}{c}{MCQ} & \multicolumn{1}{c}{MCQ} & & \multicolumn{1}{c}{MCQ} & \multicolumn{1}{c}{MCQ} & \\
 & correct & wrong & Total & correct & wrong & Total \\
\midrule
1.0 (full)    & 18 & 11 & 29 & 20 &  8 & 28 \\
0.5 (partial) &  6 &  2 &  8 &  7 &  5 & 12 \\
0.0 (none)    & 16 & 17 & 33 & 16 & 14 & 30 \\
\midrule
Total         & 40 & 30 & 70 & 43 & 27 & 70 \\
Mean score    & \multicolumn{1}{c}{0.53} & \multicolumn{1}{c}{0.40} & & \multicolumn{1}{c}{0.55} & \multicolumn{1}{c}{0.39} & \\
\bottomrule
\end{tabular}
\caption{Recognition-articulation gap: open-ended IR score conditional on MCQ correctness. ``MCQ wrong'' combines incorrect answers and ``I don't know'' responses.
Of the 40 participants who selected the correct MCQ answer pre-task, 16 (40.0\%) scored 0 on the open-ended explanation.
Post-task, 16 of 43 correct-MCQ respondents (37.2\%) could not articulate the concept.}
\label{tab:ir-gap}
\end{table*}

\subsubsection{Subjective understanding}
Indicate the extent to which you agree with the following statements. (7-point Likert scale, strongly disagree to strongly agree) \\
\begin{itemize}
    \item I know what will happen the next time I use this system, because I understand how it works.
    \item Even though I don't understand exactly how the system works, I know how to use it to get insights about the problem.
    \item It is easy to follow what the system does.
    \item I know what to do to get the advice that I need from the system, the next time I use it.
\end{itemize}

\subsubsection{NASA-TLX}
Answer the next questions on a scale from 1 (a little) to 7 (a lot)
\begin{itemize}
    \item How mentally demanding was the task?
    \item How physically demanding was the task?
    \item How hurried or rushed was the pace of the task?
    \item How successful were you in accomplishing what you were asked to do?
    \item How hard did you have to work to accomplish your level of performance?
    \item How insecure, discouraged, irritated, stressed, and annoyed were you?
\end{itemize}

\subsubsection{Qualitative questions}
This last part serves to collect general feedback. Any detail, good or bad, is useful!\\
To help: Tabs in the dashboard: (1) metric information - (2) home / weighted bias overview - (3) causal analysis - (4) extra data 
insights - (5) experiment (metric weights / group configurations) \\
Metrics: group balance, impact ratio, consistency (+licc), what-if fairness\\
\begin{itemize}
    \item Which part(s) of the dashboard helped you the most to understand the content? Why?
    \item Which part(s) of the dashboard helped you the least to understand the content? Why?
    \item Did you feel confused during the tasks? If you did, what was confusing?
    \item What could make the experience or the dashboard more accessible for you? 
    \item Any other thoughts about this, questions, or feedback?
\end{itemize}

\subsection{Appendix D: NASA-TLX subscales}
\begin{table}[h]
\centering
\small
\begin{tabular}{@{}lrrr@{}}
\toprule
Subscale & Mean & SD & Mdn \\
\midrule
Mental demand    & 6.21 & 1.06 & 7 \\
Effort           & 5.83 & 1.23 & 6 \\
Performance      & 4.51 & 1.87 & 5 \\
Frustration      & 4.29 & 1.83 & 4 \\
Physical demand  & 3.56 & 2.14 & 4 \\
Temporal demand  & 3.00 & 1.89 & 3 \\
\bottomrule
\end{tabular}
\caption{NASA-TLX subscale scores (N=70, 7-point scale).}
\label{tab:nasatlx}
\end{table}

\subsection{Appendix E: Codebook of the Semi-Structured Interviews}
\label{app:codebook}
The 8 themes and 52 codes were derived from the five semi-structured interviews of the co-design process.
The codes are grouped into
\begin{itemize}
    \item mental model and clarifications (tables \ref{tab:codebook} and \ref{tab:codebook-clarifications})
    \item agency (table \ref{tab:codebook-agency})
    \item the metric overview component (everything related to the overall score and range indication in table \ref{tab:codebook-overall-score}, general UI codes in \ref{tab:codebook-ui}, codes related to hovering and text in table \ref{tab:codebook-hover})
    \item the weighting component (table \ref{tab:codebook-weighting})
    \item metric explanations (table \ref{tab:codebook-explanations})
    \item tabs and flow (codes related to usage in table \ref{tab:codebook-usage}, and those related to placing of components in \ref{tab:codebook-placing})
    \item causal graph (table \ref{tab:codebook-causal-graph})
    \item metric/attribute/data preferences (table \ref{tab:codebook-preferences})
\end{itemize}
% -----------------------
\begin{table*}[t]
\centering
\begin{tabular}{p{0.18\textwidth}|p{0.32\textwidth}|c|p{0.32\textwidth}}
\textbf{Code} & \textbf{Description} & \textbf{\# \color{gray} (/5)} & \textbf{Proposed solution} \\
\hline
Misunderstood weighted score & The participant thought that changing weights would also change the per-metric values. One person tried to interpret the score (e.g.\ `77\% means that 77 out of 100 people get treated fairly?') instead of seeing it as a combination of metrics. & 2 & Make explicit in the intro (tutorial and/or demo video) how this score is calculated, using visuals. Show the explanation in the weighting component as well. \\
Topic difficulty & Participant understood the metrics, but found it difficult to think about what they would prioritize or the trade-offs & 2 & /\\
Need time & Mentioned that it took a lot of effort in the beginning (high initial load), but after a while a lot less. One mentioned that if using it regularly, it would not demand too much effort. & 3 & /\\
Different expectations intersections & The participant expected to see a comparison (two columns next to each other, one for each attribute), instead of intersectional data when choosing a second attribute & 1 & / (Clear once actually selected, and participant afterward added that it would not have been that useful to have a comparison there) \\
Questions difficult afterward & The questionnaire was difficult to fill in because the dashboard was no longer visible. The participants mentioned that it was clear while in the dashboard, and would be easier if they could look back, but very hard to retain that info. & 2 & / (In final study: make sure that people cannot go back to the dashboard, because they will try) \\
Misunderstood highest biases & The participant thought the component showed the search history & 1 & / (due to not reading, already present in the intro) \\
Small group: clear & Upon seeing a small group, the participant mentioned that the group is small and the results are insignificant  & 3 & / \\
Small group: not clear & Upon seeing a small group, the participant did not immediately notice that the group is small or that the results are insignificant, and started interpreting the results as usual & 2 & / (see `small group: hide results') \\
\end{tabular}
\caption{Codes related to mental model or expectations. What participants were expecting to happen, how they thought the system worked.}
\label{tab:codebook}
\end{table*}
% -----------------------
\begin{table*}[t]
\begin{tabular}{p{0.18\textwidth}|p{0.30\textwidth}|c|p{0.32\textwidth}}
\textbf{Code} & \textbf{Description} & \textbf{N (/5)} & \textbf{Proposed solution} \\
\hline
Unclear phrasing & The participant mentioned a term or sentence being difficult or unclear. & 4 & Explain or remove terms such as `intersectional' or `ratio'. Choose clearer names for the education level groups. \\
Lack of context & The participant would like a clearer context-setting in the intro, specifically that we are working towards a weighted score from these metrics. & 1 & Make this explicit in the intro (same solution as for `misunderstood weighted score'). \\
Metrics' scale difference confusing & The participant found it confusing that the metrics have different scales. & 3 & May resolve once the weighted score computation is clearer in the intro, since participants then think about the scales earlier. Clarify why statistical parity can be negative. Impact ratio changed to a percentage instead of 0-1 after the first interview. \\
Questionnaire confusion & The participant mentioned confusion around the questionnaire. & 2 & Most changes already implemented. In the matching question, the rectangle should enclose all weights instead of just the row numbers, so participants do not confuse `a' with `the first column'. Change `making decisions' to `getting insights' in the final questionnaire. \\
Optional vs.\ core component unclear & The participant wanted it to be clearer what is core functionality and what is optional. & 1 & Add `(optional)' to intro texts about group configurations and the causal graph. \\
\end{tabular}
\caption{Codes related to clarifications and phrasing. N = number of participants (out of 5) who raised the code.}
\label{tab:codebook-clarifications}
\end{table*}
% -----------------------
\begin{table*}[t]
\begin{tabular}{p{0.18\textwidth}|p{0.30\textwidth}|c|p{0.32\textwidth}}
\textbf{Code} & \textbf{Description} & \textbf{N (/5)} & \textbf{Proposed solution} \\
\hline
Would change group settings & The participant said they would adapt the group configurations to their query. & 1 & -- \\
Would leave group settings & The participant said they would use the default group settings. & 1 & -- \\
Would change weights & The participant said they would adapt the metric weights to their query. & 4 & -- \\
Would keep weights & The participant said they would use the default metric weights. & 1 & -- \\
Would use intersections & The participant said they would look at the intersectional data. & 1 & -- \\
Would not use intersections & The participant did not find the intersectional data necessary, finding it too specific. & 1 & -- \\
Misses group-specific search & The participant wants to search by group rather than only by attribute, since attribute-only still requires scrolling. & 1 & Add a group-specific search, unless it makes the interface more cluttered. \\
Misses timeline & The participant wants an overview over time (how the results have evolved). & 1 & -- (out of scope) \\
\end{tabular}
\caption{Codes related to agency: would users adapt the group configurations or weights? N = number of participants (out of 5) who raised the code.}
\label{tab:codebook-agency}
\end{table*}
% -----------------------
\begin{table*}[t]
\begin{tabular}{p{0.18\textwidth}|p{0.30\textwidth}|c|p{0.32\textwidth}}
\textbf{Code} & \textbf{Description} & \textbf{N (/5)} & \textbf{Proposed solution} \\
\hline
Overall score: lower-is-better unintuitive & Originally, the score showed how much bias there was rather than how fair the algorithm is. The participant found this unintuitive, since users typically expect a score where higher is better. & 1 & Adapted to a fairness score after the first interview. \\
Low visibility of color ranges & The participant did not initially notice the color ranges, indicating how good or bad a metric score is (2), or wanted them more visible (2). & 4 & Add a light green/orange/red background to each metric, in addition to the range indicators. \\
Did see color ranges & The participant noticed the color ranges. & 2 & -- \\
Liked color ranges & The participant found the color ranges useful. & 2 & -- \\
Add zero label & The participant requested a zero label on the range indicator bars. & 1 & Added after the first interview. \\
Keep numbers & The participant indicated that they want to see the numerical metric values in addition to the range indication and colors. & 1 & -- \\
Concerns about hiding results & The participant worried that, if the dashboard showed only the overall score and required a click for per-metric scores, people might skip the click and miss nuance. They requested keeping the per-metric scores visible. & 1 & -- \\
\end{tabular}
\caption{Codes related to the overall score and the colored range indicators. N = number of participants (out of 5) who raised the code.}
\label{tab:codebook-overall-score}
\end{table*}
% -----------------------
\begin{table*}[t]
\begin{tabular}{p{0.18\textwidth}|p{0.30\textwidth}|c|p{0.32\textwidth}}
\textbf{Code} & \textbf{Description} & \textbf{N (/5)} & \textbf{Proposed solution} \\
\hline
Crowded highest-biases component & The participant found three stacked cards crowded in the highest-biases component. They suggested either showing only the overall score with a click-through, or enlarging the cards. & 1 & -- (the participant later preferred the original layout for easier comparison between groups). \\
Low visibility of group vs.\ individual split & The participant wanted a clearer distinction between individual and group metrics. They suggested a separator line, centered titles, or grey background blocks. & 1 & Test ways to highlight the distinction without adding clutter. \\
Metric weights not visible & The participant wanted to see the weight of each metric. & 1 & Show the weight in the hover tooltip of each metric, and possibly add a weight-division bar next to the overall score, if it does not crowd the layout. \\
\texttt{meets\_requirement} bug & The participant noticed that one group rendered as `\texttt{meets\_requirement}' (the backend name) instead of `speaks English or Dutch'. & 1 & Fix the rendering of `\texttt{meets\_requirement}'. \\
Hover overlap bug & When hovering over the info icon for weights in the highest-biases component, the participant saw the group balance explanation instead. & 1 & Investigate and fix the overlapping hover area. \\
Showing insignificant results & The participant suggested hiding or removing results for small groups. One preferred showing only the group name, the participant count, and a `not enough data available' label. Another suggested a click-through to still see the results, with the extra click emphasizing that the results need caution. & 2 & Show the group with the count and a `not enough data' label. Do not offer a click-through, since the results are not significant. \\
\end{tabular}
\caption{Codes related to general UI of the metric overview. N = number of participants (out of 5) who raised the code.}
\label{tab:codebook-ui}
\end{table*}
% -----------------------
\begin{table*}[t]
\begin{tabular}{p{0.18\textwidth}|p{0.30\textwidth}|c|p{0.32\textwidth}}
\textbf{Code} & \textbf{Description} & \textbf{N (/5)} & \textbf{Proposed solution} \\
\hline
Liked hover texts & The participant said they liked the hover texts for the metric results. & 3 & -- \\
Missing overall-score hover card & The participant wanted a hover card for the overall fairness score, similar to those for the metric results. & 1 & Add a hover tooltip for the overall score. \\
Impact-ratio hover card too positive & The participant suggested adding `only' (Dutch: \emph{maar}) to the sentence, so that a low impact ratio immediately reads as bad relative to the 100\% reference group. & 1 & Add `only' / \emph{maar} to the sentence. \\
Did not see hover text & The participant did not notice that they could hover over metrics for an explanation. Upon noticing it later, they found it `very useful' and wished they had known sooner. & 1 & -- (already covered in the tutorial; other participants had no issue). \\
Confusion around 80\% rule & The participant felt confused by the 80\% rule mentioned in the impact-ratio hover text. & 2 & Remove the 80\% rule from the hover text. The color ranges already indicate whether a score is too low. \\
Missing metric-correlations explanation & The participant wanted a separate explanation, with an example, for each combination of group and metric, describing the relationships between the numbers. & 1 & -- (out of scope, too complex). \\
\end{tabular}
\caption{Codes related to the hover cards and text in the metric overview component. N = number of participants (out of 5) who raised the code.}
\label{tab:codebook-hover}
\end{table*}
% -----------------------
\begin{table*}[t]
\begin{tabular}{p{0.18\textwidth}|p{0.30\textwidth}|c|p{0.32\textwidth}}
\textbf{Code} & \textbf{Description} & \textbf{N (/5)} & \textbf{Proposed solution} \\
\hline
Missing save button & Weights are currently saved automatically. The participant felt unsure about this. & 1 & Add an explicit save button instead of auto-saving. \\
Weighting unclear without context & Seeing the component before reaching the relevant tutorial step is confusing; the purpose of the component is not self-evident. & 1 & Make the weighted score explicit in the intro. Tighten the component's text (see other UI suggestions in this codebook). \\
Sliders intuitive & The participant used the sliders without trouble. & 4 & -- \\
Pie-chart hover complaints & The participant found the pie chart's hover too slow and wanted metric explanations there too; they had not noticed the explanations lower in the component. & 1 & Speed up the pie-chart hover and include a one-sentence refresher next to the metric name. Remove the permanent metric sentences in the component, since participants did not read them. \\
Did not notice presets & The participant did not notice the preset options, although they said they would have wanted something like that. & 1 & Added presets to the tutorial after this interview. \\
Liked presets & The participant said they liked having presets. & 2 & -- \\
\end{tabular}
\caption{Codes related to the weighting component. N = number of participants (out of 5) who raised the code.}
\label{tab:codebook-weighting}
\end{table*}
% -----------------------
\begin{table*}[t]
\begin{tabular}{p{0.18\textwidth}|p{0.30\textwidth}|c|p{0.32\textwidth}}
\textbf{Code} & \textbf{Description} & \textbf{N (/5)} & \textbf{Proposed solution} \\
\hline
Hover cards too long & The participant found the metric hover cards too text-heavy. & 1 & Replace with a one-sentence refresher. \\
Confusing group balance example & The participant found the 50--50 example misleading and suggested 40 women and 60 men, so the groups do not appear to \emph{need} to be equally sized. & 2 & Use unequal group sizes in the group balance example. \\
Clear explanations & The participant said they understood all the metrics while or after reading the cards. & 3 & -- \\
Layout: missed one-liner & The participant wanted a short summary sentence so that readers who skim can still get the main point. & 1 & Make the one-sentence summary more prominent. \\
Layout: liked strengths & The participant found the strengths useful, as well as the limitations for individual fairness. & 1 & -- \\
Layout: keep clear split & The participant wanted the cards to stay clearly separated (e.g., via a selection menu). & 1 & -- \\
Liked the examples & The participant said the examples really helped their understanding. & 2 & -- \\
\end{tabular}
\caption{Codes related to the way the metrics are explained. N = number of participants (out of 5) who raised the code.}
\label{tab:codebook-explanations}
\end{table*}
% -----------------------
\begin{table*}[t]
\begin{tabular}{p{0.18\textwidth}|p{0.30\textwidth}|c|p{0.32\textwidth}}
\textbf{Code} & \textbf{Description} & \textbf{N (/5)} & \textbf{Proposed solution} \\
\hline
Usage: specific query & The participant would look for their specific queries rather than the highest biases. & 1 & -- \\
Usage: scans metrics & The participant mostly looks at per-metric values rather than the overall score, scanning for discrepancies and comparing across groups. & 1 & -- \\
Usage: overall score first & The participant first looks at the overall score, then drills into specific metrics if the score is low. & 3 & -- \\
Unclear homepage & The participant suggested a home page with the overall bias scores for high-priority attributes and a short text describing the default values and the reasoning behind them. They also suggested a separate `experiment' tab for changing weights or group configurations. & 4 & Restructure into a clearer home page (lowest fairness results plus contextual text) and a separate experiment tab. \\
No clear experiment tab & The participant wanted a single `experiment' tab containing everything needed for self-experimentation. The default view would show the consensus values; switching to the experiment tab would make the choice to deviate a conscious one. & 1 & Move the weight component into the group configuration tab and rename the tab to `experiment'. Keep the weights accessible from the metric overview through the `?'-icon. \\
\end{tabular}
\caption{Codes related to the preferred flow or expected usage of the dashboard. N = number of participants (out of 5) who raised the code.}
\label{tab:codebook-usage}
\end{table*}
% -----------------------
\begin{table*}[t]
\begin{tabular}{p{0.18\textwidth}|p{0.30\textwidth}|c|p{0.32\textwidth}}
\textbf{Code} & \textbf{Description} & \textbf{N (/5)} & \textbf{Proposed solution} \\
\hline
Keep highest biases & The participant found a list of highest biases useful, in addition to search. & 4 & -- \\
Did not find weights & The participant did not easily find the weight component. & 2 & Introduce an `experiment' tab. \\
Did not find highest biases & The participant did not easily find the highest-biases component. & 1 & Clearer home tab (see other solutions). \\
Easily found highest biases & The participant quickly found the highest-biases component. & 2 & -- \\
Knows what, not where & The participant could describe the weighted overview cards but could not locate them in the dashboard. & 1 & Improve flow clarity (see other solutions in this section). \\
Difficulty finding specific group & While searching for a specific group, the participant did not go directly to the correct component, instead scrolling or navigating the menu until stumbling on it (sometimes quickly, sometimes slowly). & 4 & Improve flow clarity (see other solutions in this section). \\
\end{tabular}
\caption{Codes related to what should be kept, and how easily components are found. N = number of participants (out of 5) who raised the code.}
\label{tab:codebook-placing}
\end{table*}
% -----------------------
\begin{table*}[t]
\begin{tabular}{p{0.18\textwidth}|p{0.30\textwidth}|c|p{0.32\textwidth}}
\textbf{Code} & \textbf{Description} & \textbf{N (/3)} & \textbf{Proposed solution} \\
\hline
CG: actionable & The participant liked the component and would use it when looking for solutions. & 1 & -- \\
CG: difficult & The participant would prefer a table with one row per correlation; the graph was difficult to read. & 1 & -- (this participant had earlier noted that tables were difficult for them, so the preference is not generalizable). \\
CG: clear & The participant found the graph intuitive to read. & 1 & -- \\
\end{tabular}
\caption{Codes related to the causal graph. In the exploratory interviews, we did not push participants toward the causal graph: the tutorial mentioned the tab but did not direct them there. Participants who wanted to could open it. These codes capture whether the graph was clear to them and whether they found it useful.
N = number of participants (out of 3 who opened the graph) who raised the code.}
\label{tab:codebook-causal-graph}
\end{table*}
% -----------------------
\begin{table*}[t]
\begin{tabular}{p{0.18\textwidth}|p{0.30\textwidth}|c|p{0.32\textwidth}}
\textbf{Code} & \textbf{Description} & \textbf{N (/5)} & \textbf{Proposed solution} \\
\hline
Prefers consistency & The participant preferred consistency. & 1 & -- \\
Intuitiveness of metrics & One participant found group balance and what-if fairness more intuitive than the other two metrics; another found consistency the most intuitive. & 2 & -- \\
Prefers impact ratio & The participant liked impact ratio, saying it made the results `more tangible'. & 1 & -- \\
Difficult to define similarities & The participant noted the difficulty of defining what `similar' candidates means for the individual-level metrics. & 1 & -- \\
Important attribute & The participant called out a single attribute as the most important: age (1$\times$), nationality (1$\times$), and language (1$\times$). & 3 & -- \\
Wrong granularity & The participant wanted results split by sector and by blue-collar vs.\ white-collar rather than per individual job, noting that this information is always available on vacancies. & 1 & Add the blue-collar/white-collar split. \\
\end{tabular}
\caption{Codes that captured explicit metric, attribute, or data preferences. N = number of participants (out of 5) who raised the code.}
\label{tab:codebook-preferences}
\end{table*}

\end{document}